\documentclass[reprint,aps,prb,twocolumn,groupedaddress,amsmath,amssymb,superscriptaddress,floatfix]{revtex4-1}
\usepackage{graphicx}
\usepackage{dcolumn}
\usepackage{bm}
\usepackage{verbatim}
\usepackage[caption=false]{subfig}
\usepackage{url}
\usepackage{listings}
\usepackage[pdftex,colorlinks=true,linkcolor=blue,citecolor=blue,urlcolor=blue]{hyperref}

\begin{document}


\title{Scalable and fast heterogeneous molecular simulation with predictive parallelization schemes}

\author{Horacio V. Guzman}
\email{vargas@mpip-mainz.mpg.de}
\affiliation{Max Planck Institute for Polymer Research, Ackermannweg 10, 55128 Mainz, Germany}
\author{Christoph Junghans}
\affiliation{Computer, Computational, and
Statistical Sciences Division, Los Alamos National Laboratory, Los Alamos, NM 87545, USA}
\author{Kurt Kremer}
\affiliation{Max Planck Institute for Polymer Research, Ackermannweg 10, 55128 Mainz, Germany}
\author{Torsten Stuehn}
\affiliation{Max Planck Institute for Polymer Research, Ackermannweg 10, 55128 Mainz, Germany}

\date{\today}

\begin{abstract}
Multiscale and inhomogeneous molecular systems are challenging topics in the field of molecular simulation. In particular, modeling biological systems in the context of multiscale simulations and exploring material properties are driving a permanent development of new simulation methods and optimization algorithms. In computational terms, those methods require parallelization schemes that make a productive use of computational resources for each simulation and from its genesis. Here, we introduce the heterogeneous domain decomposition approach which is a combination of an heterogeneity sensitive spatial domain decomposition with an \textit{a priori} rearrangement of subdomain-walls. Within this approach, the theoretical modeling and scaling-laws for the force computation time are proposed and studied as a function of  the number of particles and the spatial resolution ratio. We also show the new approach capabilities, by comparing it to both static domain decomposition algorithms and dynamic load balancing schemes. Specifically, two representative molecular systems have been simulated and compared to the heterogeneous domain decomposition proposed in this work. These two systems comprise an adaptive resolution simulation of a biomolecule solvated in water and a phase separated binary Lennard-Jones fluid.

\begin{description}
\item[PACS numbers]
tbd.
\end{description}
\end{abstract}

\pacs{Valid PACS appear here}
\maketitle


\section{\label{intro}Introduction}

Molecular simulation has proven being of vital importance for driving the theory and the interpretation of experiments in diverse research fields, including soft matter research~\cite{KK-FMP2002,attig04,voth08,holm0509,peter_faraday_2010,Halverson2013,OlveraSM2k15}. In particular, molecular dynamics (MD) methods have facilitated the study of systems consisting of millions of particles~\cite{RapaportCPC1991} and timescales up to milliseconds~\cite{ShawPCHPCNSA2009}. These milestones in MD simulations have been achieved by promoting two facts: the intrinsic highly parallel environment enabled by the MD algorithm and the development of new optimized algorithms for several MD systems~\cite{KroegerP2013,RapaportJPCM2014,ShawCOSB2014,FieldABB2015,HolzerP2017}. Out of equilibrium (inhomogeneous) and multiscale simulations constitute recent milestones in molecular simulations of soft matter which have driven in the past few years the development of new methods aiming to provide a physically consistent treatment, highly accurate and computationally efficient molecular modeling. In particular, concurrent multiscale modeling, mapping atomic configurations to coarse grained models and partitioning the system in different resolution regions~\cite{praprotnik_multiscale_2008,delgado08a,delgado-buscalioni_coupling_2009,Donadio_PRL_2013-hadres_molliq}. As well as several enhancements developed for studying inhomogeneous systems; like phase separated ones~\cite{SmrekPRL2017}, crystal growth~\cite{RaduPRL2017}, biomembranes~\cite{BereauM2015,PluhackovaBM2015} and proteins~\cite{ShawPCHPCNSA2009,ShawCOSB2014}.

Understanding soft matter at multiple time and length scales poses particular computational demands where microscopic physical phenomena strongly couple to macroscopic representations e.g. by adjusting resolutions on the fly between models of multiple costs. Consequently, suited parallelization schemes should keep these considerations in mind and meet both requirements: scalability and simulation speed. Furthermore available load-balancing algorithms are based on measurements of computational runtime, although they are limited for predicting the domain decomposition of an inhomogeneous simulation from its initial configuration. A similar situation arises when tackling out-of-equilibrium molecular systems with prescribed levels of heterogeneity, which are commonly given as characteristics and topology of the different particles within the simulation setup. 

In this article we introduce the Heterogeneous Spatial Domain Decomposition Algorithm (HeSpaDDA). This algorithm is not based in any runtime measurements, it is rather a predictive computational resources allocation scheme. The HeSpaDDA algorithm is a combination of an heterogeneity (resolution or spatial density) sensitive processor allocation with an initial rearrangement of subdomain-walls. The initial rearrangement of subdomain-walls within HeSpaDDA anticipates favorable cells distribution along the processors per simulation box axis by moving cell boundaries according to the resolution of the tackled region in the molecular system. In a nutshell, the proposed algorithm will make use of \textit{a priori} knowledge of the system setup. Specifically, the region which is computationally less expensive. This inherent load-imbalance could come from different resolutions or different densities. The algorithm will then propose non-uniform domain layout, \emph{i.e.} domains of different size and its distribution amongst compute instances. This could lead to significant speedups for systems of the aforementioned type over standard algorithms,\emph{e.g.} spatial Domain Decomposition (DD)~\cite{lammps95} or spatial and force based DD~\cite{de_shaw}.

The multiscale simulation method we have chosen to validate the capabilities of HeSpaDDA is the Adaptive Resolution Scheme (AdResS). In particular, for a dual resolution simulation like AdResS we tackle an expensive model in the high-resolution region and a less-expensive model elsewhere. In computational terms, this means that some processors have more work to do than others at the initial domain-decomposition of a dual resolution simulation. AdResS methods have been employed for the concurrent simulation of diverse length scale systems interfacing different resolutions of simulation techniques, ranging from concurrent simulations of classical atomistic and coarse-grained models~\cite{praprotnik_adaptive_2005,praprotnik_adaptive_2007-1,praprotnik_corrigendum:_2009,Potestio2013b,Kreis2014,Kremer_JCP_2015-adresprot,kreisEPJST2015,Praprotnik_JCTC_2015-dna}, to coupling classical atomistic and path-integral models~\cite{Poma2010,Poma2011,KreisJCTC2k16}, as well as, interfacing particle based simulations with continuum mechanics~\cite{delgado08a,delgado-buscalioni_coupling_2009}.

\begin{figure}[!t]
\centering
\includegraphics[clip,width=0.6\columnwidth,keepaspectratio]{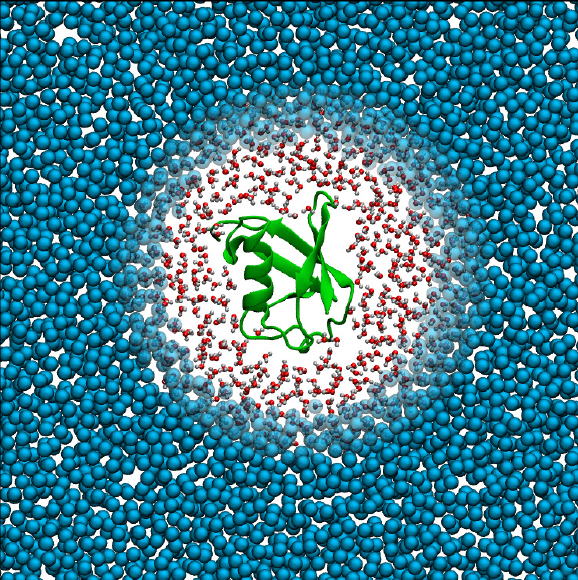}
\caption{AdResS simulation of an atomistic protein and its atomistic hydration shell, coupled to a coarse-grained particle reservoir via a transition region~\cite{Kremer_JCP_2015-adresprot}.}
\label{fig: adress app}
\end{figure}

The bimolecular AdResS simulation (see Figure~\ref{fig: adress app}) tackled with HeSpaDDA, is a cubic simulation system with a spherical high-resolution region, comprised by an aqueous solution of the regulatory protein ubiquitin~\cite{Kremer_JCP_2015-adresprot}. Understanding biomolecular function is fundamental to a broad variety of research domains, including drug discovery, food processing and bioprocess engineering, to name but a few. In recent years, the AdResS approach has been successfully applied to an increasing number of biomolecular systems~\cite{Kremer_JCP_2015-adresprot,Praprotnik_JCTC_2015-dna,Praprotnik_JChemPhys_2014-adres_bio}, with the goal of obtaining both computational speedup and additional insight in the systems properties.

\begin{figure}[]
\centering
\includegraphics[clip,width=0.9\columnwidth,keepaspectratio]{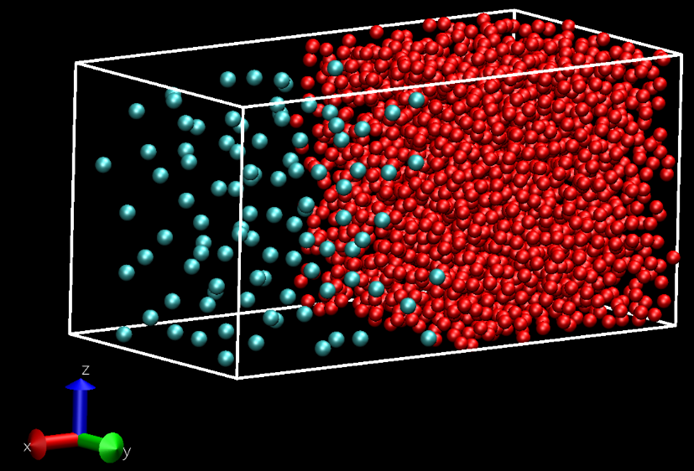}
\caption{Lennard-Jones system of a binary fluid separated by two phases of equal volume but different $\sigma$ (6 to 1).}
\label{fig: LJ6to1Basic}
\end{figure}

In the case of inhomogeneous single-scaled systems we have tackled the system described in Figure~\ref{fig: LJ6to1Basic}, which contains two phase separated fluids of Lennard-Jones particles presenting a 6 to 1 $\sigma$ ratio of heterogeneity.

The capabilities of the HeSpaDDA algorithm have been benchmarked by using two MD simulation packages. The AdResS simulations have been carried out by employing the ESPResSo++ MD package (static domain decomposition). While for the single-scale phase separated 6 to 1 Lennard-Jones binary fluid has been simulated with the GROMACS MD package and makes use of its embedded dynamic load balancing features. However the HeSpaDDA algorithm can be implemented within any other MD packages supporting multiscale and$/$or inhomogeneous simulations.

\section{\label{adress}Adaptive resolution scheme}

\begin{figure}[ht!]
\centering
\includegraphics[clip,width=0.7\columnwidth,keepaspectratio]{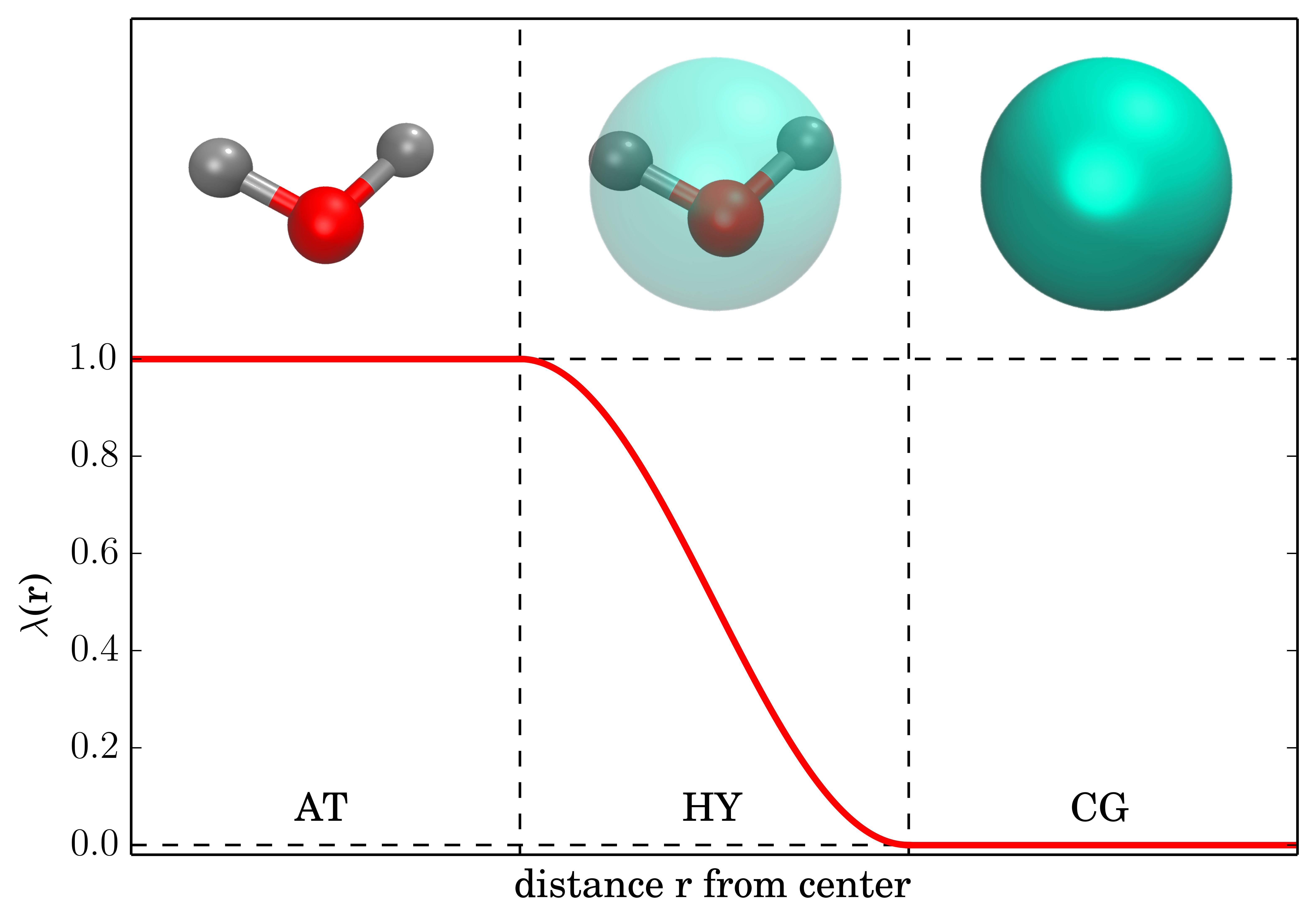}
\caption{Schematic of the resolution function $\lambda(\mathbf{R}_\alpha)$ used for the interpolation between the AT and CG forcefields in the adaptive resolution methodology. In the presented case, an atomistic water model is coupled with a coarse-grained one-particle-per-molecule description in the CG region.}
\label{fig: adress scheme}
\end{figure}

In the Adaptive Resolution Simulation Scheme (AdResS), a region in space is defined in which molecules are modeled using atomistic detail, while coarse-grained models are used elsewhere (see Figure~\ref{fig: adress app}). Particles can freely diffuse between atomistic and coarse-grained regions, smoothly changing their resolution as they cross a hybrid or transition region (see Figure~\ref{fig: adress scheme}). 

Here, we review the theoretical basis behind the adaptive  resolution methodology. A typically small part of the system, the atomistic (AT) region, is described on the atomistic level and coupled via a hybrid (HY) transition region, to the coarse-grained (CG) region, where a coarser, computationally more efficient model is used. The interpolation is achieved via a resolution function $\lambda({\mathbf{R_\alpha}})$, a smooth function of the center of mass position $\mathbf{R_\alpha}$ of molecule $\alpha$. For each molecule, its instantaneous resolution value $\lambda_\alpha = \lambda({\mathbf{R_\alpha}})$ is calculated based on the distance of the molecule from the center of the AT region. It is 1 if the molecule resides within the AT region and smoothly changes via the HY region to 0 in the CG region (see Figure \ref{fig: adress scheme}).

The interpolation between atomistic and coarse-grained non-bonded forcefields can be performed at the level of energies (Hamiltonian-AdResS) or of forces (Force-AdResS), as explained in the following section for this publication we employ the Force-AdResS method.

Note that, in addition to the non-bonded interactions, bonded potentials are also usually present. In typical AdResS applications they are not subject to interpolation. As these are computationally significantly easier to evaluate, they do not play a role in the parallelization scheme presented here.
%

\subsection{\label{sec:level2}Force interpolation}
In the force interpolation scheme, the original AdResS technique~\cite{Kremer_JCP_2005-onthefly,Praprotnik2008review}, two different non-bonded force fields are coupled as 
\begin{equation}\label{eq:adress}
\mathbf{F}_{\alpha|\beta} = \lambda(\mathbf{R_\alpha})\lambda(\mathbf{R_\beta})\mathbf{F}_{\alpha|\beta}^{\text{AT}} + \left(1-\lambda(\mathbf{R_\alpha})\lambda(\mathbf{R_\beta})\right)\mathbf{F}_{\alpha|\beta}^{\text{CG}},
\end{equation}
where $\mathbf{F}_{\alpha|\beta}$ is the total force between the molecules $\alpha$ and $\beta$ and $\mathbf{F}_{\alpha|\beta}^{\text{AT}}$ defines the atomistic force-field which is decomposed into atomistic forces between the individual atoms of the molecules $\alpha$ and $\beta$. Finally, $\mathbf{F}_{\alpha|\beta}^{\text{CG}}$ is the coarse-grained force between the molecules, typically evaluated between their centers of mass.

\section{Heterogeneous Spatial Domain Decomposition Algorithm}
\label{mrDD}

Molecular simulations in soft matter investigations, such as polymer and biomolecular research require the evaluation of efficient parallelization schemes \emph{e.g.} algorithms~\cite{KroegerP2013,ShawCOSB2014,HolzerP2017,lammps95,gromacs4,de_shaw}. In particular, initial domain decomposition algorithms can be based on the distribution of atoms, forces or the space in a given simulation domain, and also a combination of both forces and space~\cite{de_shaw}. After a certain amount of simulation time-steps, Dynamic Load Balancing (DLB) schemes can be activated, like the work-sharing~\cite{FlynnDLB1992,HummelSC1991,PolyITC1987} and work-stealing~\cite{BlumofeICS1994, KruskalPP1985,RudolphSPAA1991} ones. In general, distributed runtime systems with High Performance Computing (HPC) focus are employed to achieve DLB, such as: Charm++~\cite{CharmPP}, HPX~\cite{HPX}, Legion~\cite{Legion}, among others. However, initial domain decomposition algorithms are not aware of the inherent region-based distribution of multiscale and/or inhomogeneous systems from the simulation setup. The idea behind an initial and predictive DD algorithm is to reach higher scalability and efficiency by assuming a steady runtime behavior of the simulation platform and hence deliver the domain decomposition within the simulation setup and without any initial overhead in terms of load per processor. Naturally, the proposed algorithms are complementary to the dynamic load balancing ones that will take care of the platform dependent imbalance along extensive production runs. Interestingly, the trigger for such DLB algorithms could be extended by using HeSpaDDA, e.g. for the Lennard-Jones binary mixture described in this article, DLB could be required only after 100 thousand MD-steps. Whereby the common trigger used per default in DLB algorithms is in the range from 1 to 100 MD-steps. 

\begin{figure}[!t]
\centering
\includegraphics[clip,width=0.9\columnwidth,keepaspectratio]{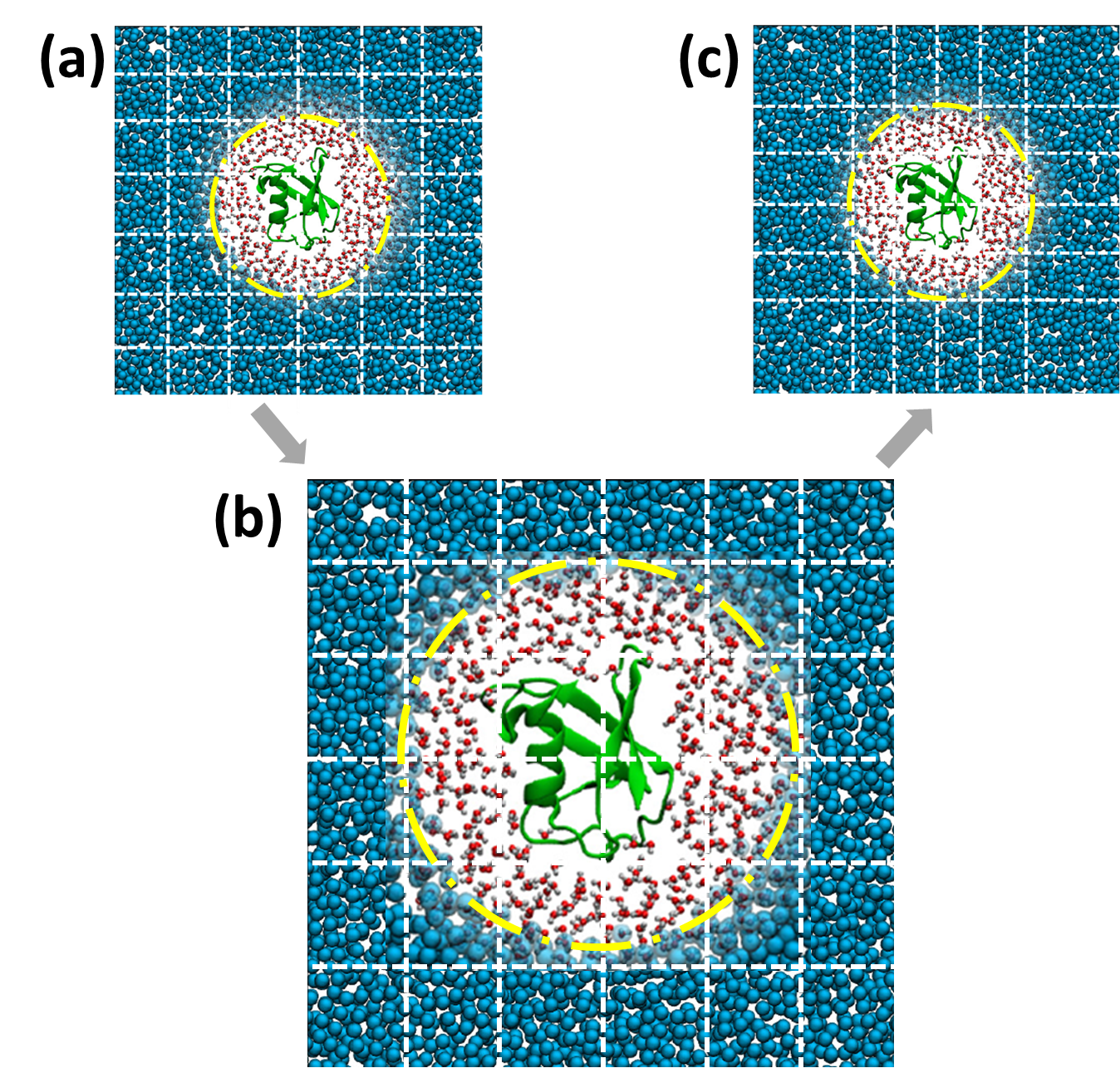}
\caption{AdResS simulation of an atomistic protein, its atomistic hydration shells and CG water particles. (a) The dashed lines illustrate the spatial domain decomposition (sDD) of such simulation; (b) shows the intermediate state where only the high-resolution region (dash-dotted circle in yellow) has been spatially rescaled in order to reach the same resolution as the low-resolution one, then in (c) the processors are allocated according to the Heterogeneous Spatial Domain Decomposition Algorithm (HeSpaDDA) by distinguishing the resolution region type. Note that in (c) no more spatially rescaling is shown and the initial system configuration is reached. In all cases the total grid is 6x6 processors, however the processors allocation in the high-resolution region ($P_{HR}=P^x_{HR}*P^y_{HR}$) changes from (a) to (c), namely, from 2x3 to 4x4 subdomains. Due to illustrative reasons only two dimensions are depicted.}
\label{fig: DRD_idea}
\end{figure}

Here, we introduce Heterogeneous Spatial Domain Decomposition Algorithm (HeSpaDDA) that is a combination of an heterogeneity (resolution or density) sensitive spatial domain decomposition (sDD) with an initial subdomain-walls rearrangement. The fact that the simulation setup presents a degree of heterogeneity along one of the simulation box axes, like the AdResS (resolution heterogeneous) or Lennard-Jones binary mixture ones (density heterogeneous), allow us to identify the simulation box regions with high and low resolutions and hence assign computational resources according to the resolution and volume ratios between those regions from the very beginning of the production run (see Figure~\ref{fig: DRD_idea}).\\

HeSpaDDA can be divided in two main modules: the first one is devoted to the allocation of processors along the simulation box axes and the second one to the cells partitioning of resolution dependent volumes to processors subdomains. Here each cell is defined as $r_{cell} \leq r_c+r_s$ with $r_c$ and $r_s$ as the cut-off and skin lengths respectively~\cite{allen87}. A prevailing reference value used in the modeling is $r_{cell}=r_c+r_s$~\cite{Halverson2013}. The first module is of crucial importance because the processor allocation defines the productive scaling of the system to be simulated already at the initial configuration of the simulation run. The development of HeSpaDDA is illustrated in Figure~\ref{fig: DRD_idea} where we observe how the subdomain-walls rearrangement is performed. First the processor allocation as for sDD are shown (Figure~\ref{fig: DRD_idea}a), then the high-resolution region (inside the dash-dotted circle in yellow) has been temporarily rescaled only for processor allocation special-purpose (Figure~\ref{fig: DRD_idea}b). Once the processor allocation has been set, the processors are adjusted to the regions according to the low or high resolution type (Figure~\ref{fig: DRD_idea}c). Moreover, HeSpaDDA includes two types of system configurations \textit{i.e.} cubic and noncubic setups. From here on, in this paper the setup with spherical high resolution region corresponds to a cubic case (Figure~\ref{fig: adress app}) and the slab-like to the noncubic one (Figure~\ref{fig: LJ6to1Basic}). While cubic configurations tend to present the same amount of processors per simulation box axes \textit{i.e.} $P_x=P_y=P_z$ to achieve efficient scaling, the noncubic configuration requires an additional optimization step in order to define the best triplet $(P_x,P_y,P_z)$. In particular, for the system described in Section~\ref{1to6Section} $P_x$  is independent from $P_y$ and $P_z$ (with $P_y=P_z$). Note here that HeSpaDDA is able to handle any box axes configuration required by the system setup (flow chart shown in the Appendix~\ref{appA}).

In the second module, the algorithm receives the number of processors per region as an input and starts distributing cells along box axes in each subdomain according to the resolution type. Furthermore, this module controls the size and cubicity of all cells (see Section~\ref{algoDes}). A successful cell's cubicity check guarantees that the communication between surfaces of different neighbors-cells of the parallelepiped are or tend to be equal in each box axes $X$, $Y$ and $Z$. Interestingly this holds for homogeneous and inhomogeneous systems where HeSpaDDA is applied. The intrinsic discrepancies between low and high resolution regions are considered in the method by a differentiated distribution of cells according to the type of resolution in each region. For more details on the algorithm flow chart see the Appendix~\ref{appA}.

The validation of the HeSpaDDA algorithm has been performed with two archetypical heterogeneous systems, a biomolecule solvated in water (Force-AdResS) and a phase separated 6 to 1 Lennard-Jones binary fluid. Both systems represent simulation setups, where the combination of this initial domain decomposition algorithm and dynamic load balancers is expected to increase productivity along an extensive production run (as verified in Section~\ref{1to6Results}). Likewise, for slowly varying systems (within this article referred to simulations demanding Verlet-list rebuilds with periodicity higher than 7 $steps_{MD}$) the HeSpaDDA algorithm can also be used to adapt the cell-grid of concurrent multiscale simulations based on performance indicators, such as the number of interactions or particle density per processor.
  
\subsection{Modeling and scaling laws for HeSpaDDA}
\label{modeling}
In order to circumvent the inclusion of platform dependent run times we based the scaling law for the computation time of HeSpaDDA on the approach proposed in Reference~\cite{lammps95} (see Equation~\ref{eq:scaling}). This approach considers the computation time $t$ in the context of a homogeneous and cubic system by considering an MD integrator in accordance to the velocity Verlet and Linked-Cell-List (LCL) algorithm. In Equation~\ref{eq:scaling}, the first term reflects pure force calculation time assuming that computed interactions are short-range. Moreover, the second term tackles the particle positions exchange between adjacent-processors in three dimensions. The particle positions need to be updated in every time step and hence it is part of the computation time $t$. Leading to,
\begin{equation}\label{eq:scaling}
t \propto \frac{N_{AT}}{2P}+6r_{cell}k_{hw}\left(\frac{N_{AT}}{P}\right)^{2/3}
\end{equation}
Where $N_{AT}$ is the total number of atomistic particles, $P$ is the total number of processors, $r_{cell}$ is the sum of the interaction cutoff radius and skin length and $k_{hw}$ is a constant describing the communication taking place in each force calculation with strong depence on the underlying hardware platform. From here on this constant will be defined as $k_{hw}\equiv1$, emphasizying the exclusion of platform dependent run times.      

Data communication modeling in high-performance computing is a subject of constant development~\cite{OspreyHPCS2014,MalikCCPE2016}. In particular for message-passing parallel programming models where blocking, as well as, nonblocking communications while synchronizing nearest-neighbor processes affect the performance of the simulations~\cite{HenriPRE2015}.

Taking the first term of Equation~\ref{eq:scaling} as a starting point, we consider two resolutions within the same simulation box and hence the number of processors located in both resolutions, as well. The number of particles in each resolution region depends on the volume ratio of each resolution region and the total volume $V$. Having for the low-resolution (LR) region,
\begin{equation}\label{eq:N_LR}
N_{LR}=\frac{V_{LR}}{V}N
\end{equation}
where $N_{LR}$ is the number of particles in the low-resolution region and $\frac{V_{LR}}{V}$ is the ratio of volumes between the low-resolution region and the total simulation box. Note that $N$ is the total number of particles represented at the lowest-resolution. Mathematically $N$ is defined as $N=N_{LR}+N_{HR}/R_{SH}^{res}$ with \emph{spatial heterogeneity resolution ratio} $R_{SH}^{res}$ presented in Equation~\ref{eq:R_MS}. Furthermore, the number of particles in the high-resolution (HR) region are given by,
\begin{equation}\label{eq:N_HR}
N_{HR}=R_{SH}^{res}\frac{V_{HR}}{V}N
\end{equation}
with,
\begin{equation}\label{eq:R_MS}
R^{res}_{SH}=\frac{N^{res}_{HR}}{N^{res}_{LR}}
\end{equation}
where $N^{res}_{HR}$ are the number of entities in the high-resolution region that correspond to one entity in the low-resolution one $N^{res}_{LR}$. Mapping the atomistic water molecule to the coarse-grained model (as depicted in Figure~\ref{fig: adress scheme}), results in $R_{SH}^{res}=3$. The ratio of volumes between the high-resolution region and the total simulation box is $\frac{V_{HR}}{V}$ comprised in Equation~\ref{eq:N_HR}.

From the viewpoint of the total number of processors, a straightforward spatial domain decomposition approach would decompose the system by taking into account the volume ratio of the different resolutions and the total simulation box. Letting the number of processors allocated in the low-resolution region as,

\begin{equation}\label{eq:P_LR}
P_{LR}=\frac{V_{LR}}{V}P
\end{equation}
Under such perspective, the processors allocated in the high-resolution region for sDD  are given by,
\begin{equation}\label{eq:P_HR}
P_{HR}=\frac{V_{HR}}{V}P
\end{equation}
where $\frac{V_{HR}}{V}$ is the volume ratio of the high-resolution region and the total simulation box. The modeling given by Equations~\ref{eq:N_LR},~\ref{eq:N_HR},~\ref{eq:P_LR} and ~\ref{eq:P_HR}, can be written as an extension to Equation~\ref{eq:scaling},
\begin{equation}\label{eq:sDD_scaling}
\begin{split}
t_{sDD} & \propto \frac{1}{2}\left(\frac{N_{LR}}{P_{LR}}+\frac{N_{HR}}{P_{HR}}\right) \\ 
& +6r_{cell}k_{hw}\left(\frac{N_{LR}}{P_{LR}}+\frac{N_{HR}}{P_{HR}}\right)^{2/3}
\end{split}
\end{equation}

The term $\left(\frac{N_{LR}}{P_{LR}}+\frac{N_{HR}}{P_{HR}}\right)^{2/3}$ due to particles communication also includes the coupling term of the communication between the high and low resolution regions. Moreover, the coupling or resolution transition effect of the pure force calculation of low and high resolutions is included within the term $\left(\frac{N_{LR}}{P_{LR}}+\frac{N_{HR}}{P_{HR}}\right)$ of Equation~\ref{eq:sDD_scaling}. Following the AdResS method described in Section~\ref{adress} the resolution transition effect would be reflected in the hybrid region (see Figure~\ref{fig: adress scheme}).

The simplest and straightforward model for the spatial allocation of particles per processor is reflected in Equation~\ref{eq:sDD_scaling} with the volume resolution ratios shown in Equations~\ref{eq:N_LR},~\ref{eq:N_HR},~\ref{eq:P_LR} and ~\ref{eq:P_HR}. However for heterogeneous simulations, such an approach has two drawbacks: there is an imbalanced distribution of particles per processor from the beginning of the simulation and the communication overhead is not minimized for the heterogeneous transition regions~\cite{FlynnDLB1992}. The results of employing the model given by Equation~\ref{eq:sDD_scaling} could hinder both the scalability and speed of the tackled heterogeneous simulation (as demonstrated in Section~\ref{results}).
The development of HeSpaDDA's model is based on a spatial distribution of particles of heterogeneous systems which fulfills the criteria given in Equation~\ref{eq:parts-to-procs}. By substituting Equations~\ref{eq:N_LR} and ~\ref{eq:N_HR} with $P_{LR}=P-P_{HR}$ a new Equation for $P_{HR}$ has been found (see Equation~\ref{eq:P_HR_DRD}). Consequently, HeSpaDDA modeling treats both low and high resolution regions as equals in terms of the allocation of particles per processor.
\begin{equation}\label{eq:parts-to-procs}
\frac{N_{HR}}{P_{HR}}=\frac{N_{LR}}{P_{LR}}
\end{equation}
To this end, Equations~\ref{eq:P_LR} and ~\ref{eq:P_HR} turn into Equations~\ref{eq:P_LR_DRD} and ~\ref{eq:P_HR_DRD}. Interestingly the proposed heterogeneous domain decomposition method is sensitive to both the region size and the heterogeneity types (resolution) for the processor allocation. Such modeling optimizes the initial or static distribution of load and hence minimizes communication overhead. In this new context the number of processors distributed in the low-resolution region is defined as,
\begin{equation}\label{eq:P_LR_DRD}
P_{LR}^{hDD}=P-P_{HR}^{hDD}
\end{equation}
where $P_{HR}^{hDD}$ are the number of processors distributed in the high-resolution region according to the HeSpaDDA algorithm. From now on, the nomenclature used in the Equations with superscript \emph{hDD} correspond to the HeSpaDDA method. By replacing Equations~\ref{eq:N_LR},~\ref{eq:N_HR} and~\ref{eq:P_LR_DRD} into~\ref{eq:parts-to-procs}. Leads to,
\begin{equation}\label{eq:P_HR_DRD}
P_{HR}^{hDD}=\frac{R_{SH}^{res}V_{HR}}{V+V_{HR}(R_{SH}^{res}-1)}P
\end{equation}
The concept behind the coefficient $R_{SH}^{res}$ in Equation~\ref{eq:P_HR_DRD} is to emulate an increase of the high-resolution volume by rescaling it to low-resolution units as shown in Figure~\ref{fig: DRD_idea}b. Moreover, to be consistent with such high-resolution region increment, the denominator of Equation~\ref{eq:P_HR_DRD} has been also expanded to $V+V_{HR}(R_{SH}^{res}-1)$. Note here that $V$ already contains one volume $V_{HR}$ and hence the denominator coefficient is $(R_{SH}^{res}-1)$. 
Here, we remark that neither the simulation parameters $V_{HR}$ nor $V$ are scaled or increased for the simulation run. Nonetheless, HeSpaDDA employs the volume rescaling for achieving a balanced processor allocation among the different spatial resolution regions. In light of the limitation of allocating processors per simulation box axes at the simulation start, the number of processors will be rounded up to the closest integer for $P_{HR}^{hDD}$ and $P_{LR}^{hDD}$ is calculated from Equation~\ref{eq:P_LR_DRD}.

\begin{figure}[!b]
\centering
\includegraphics[clip,width=0.8
\columnwidth,keepaspectratio]{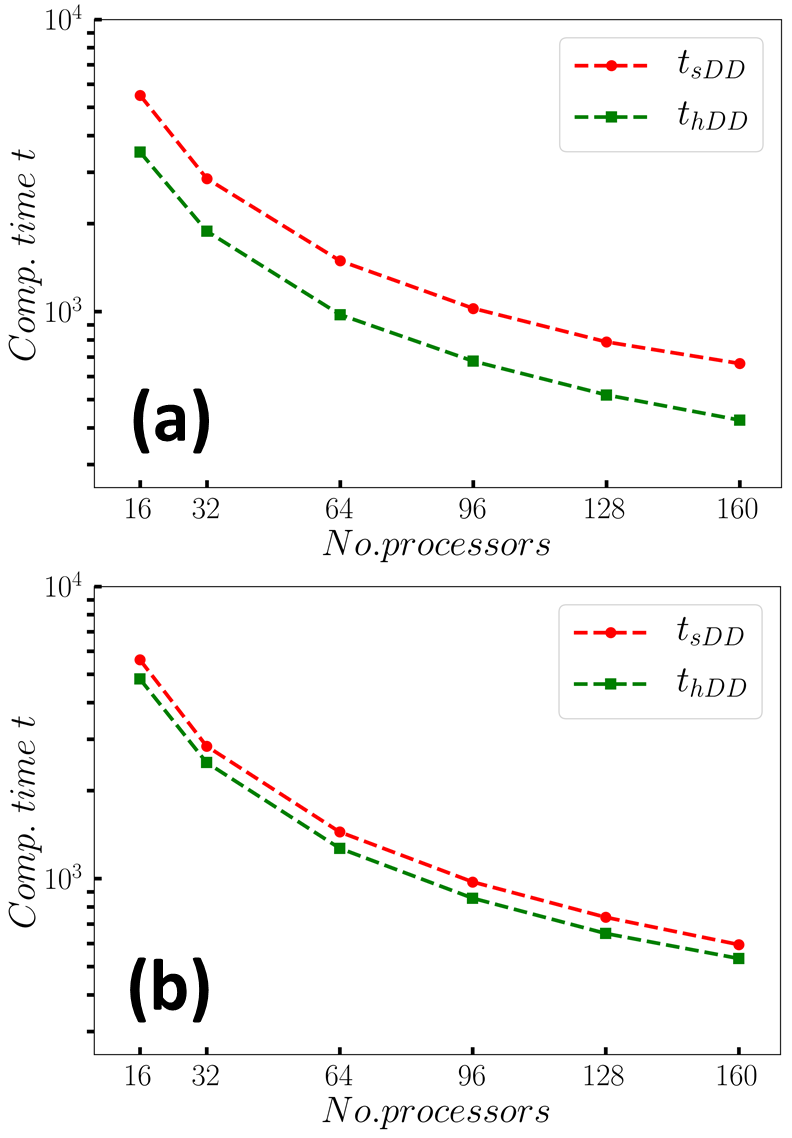}
\caption{Theoretical calculation of the computational time for the spatial domain-decomposition algorithm (see Equation~\ref{eq:sDD_scaling}) and the HeSpaDDA one (see Equation~\ref{eq:DRD_scaling}), as a function of a variable number of processors $P$. The calculation parameters for the multiscale and cubic system (a) are: $r_{c}=1.0$, $r_s=0.3$, $r_{cell}=r_{c}+r_s$, $R_{SH}^{res}=3$, $N=38084+(76916/R_{SH}^{res})$, $V=1157.625\sigma^3$, $V_{HR}=140.608\sigma^3$ and $V_{LR}=V-V_{HR}$. The calculation parameters for the inhomogeneous and noncubic system (b) are: $r_{c}=0.9$, $r_{cell}=r_{c}$, $R_{SH}^{res}=6$, $N=20828+(124964/R_{SH}^{res})$, $V=8192\sigma^3$, $V_{HR}=4096\sigma^3$ and $V_{LR}=V-V_{HR}$. All in reduced units with $\sigma=\epsilon=m=1$.}
\label{fig:DRD_sDD_theory}
\end{figure}

Thus, the time scaling law for the HeSpaDDA is,
\begin{equation}\label{eq:DRD_scaling}
\begin{split}
t_{hDD} & \propto \frac{1}{2}\left(\frac{N_{LR}}{P_{LR}^{hDD}}+\frac{N_{HR}}{P_{HR}^{hDD}}\right) \\
& +6r_{cell}k_{hw}\left(\frac{N_{LR}}{P_{LR}^{hDD}}+\frac{N_{HR}}{P_{HR}^{hDD}}\right)^{2/3}
\end{split}
\end{equation}
In Figure~\ref{fig:DRD_sDD_theory}a theoretical comparison between the scaling laws of Equations~\ref{eq:sDD_scaling} and~\ref{eq:DRD_scaling} has been carried out. Whereby the new algorithm (HeSpaDDA) improves the performance of the former spatial domain decomposition for two archetypical heterogeneous systems. From the viewpoint of this theoretical framework, the signatures of runtime based ``dynamic'' imbalance is disregarded. Certainly the modeling of the here presented heterogeneous algorithm is static. Nevertheless in the simulation code the method can be combined to external dynamic load balancing schemes (shown in Section~\ref{results}).
%
The inhomogeneous system tackled in Figure~\ref{fig:DRD_sDD_theory}a is similar in terms of simulation parameters to the ubiquitin solvated in water (see Section~\ref{ubiquitin}). However in this theoretical example we considered pure atomistic water in the high-resolution region (no ubiquitin in the high-resolution region). While Figure~\ref{fig:DRD_sDD_theory}b resembles the phase separated binary Lennard-Jones fluid. Calculating the corresponding computational times for \emph{sDD} (Equation~\ref{eq:sDD_scaling}) and \emph{hDD} (Equation~\ref{eq:DRD_scaling}) as a function of $P$ we observed in all studied cases improved timing for HeSpaDDA. 

\begin{figure}[!ht]
\centering
\includegraphics[clip,width=0.97\columnwidth,keepaspectratio]{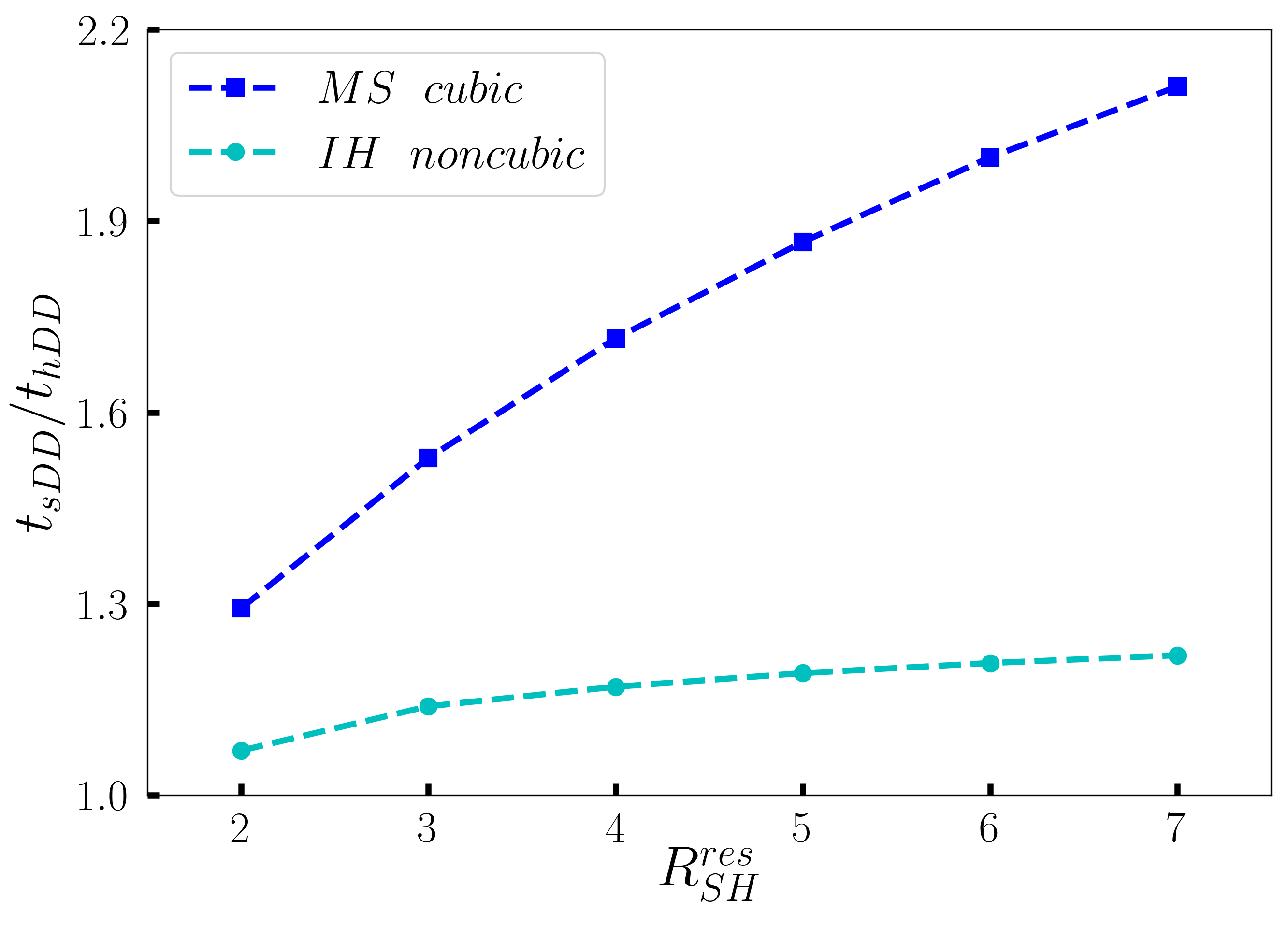}
\caption{Theoretical calculation of the computational time for the spatial domain-decomposition algorithm (see Equation~\ref{eq:sDD_scaling}) over the HeSpaDDA one (see Equation~\ref{eq:DRD_scaling}), as a function of variables (a) $R_{SH}^{res}$ with the simulation parameters: $R_{SH}^{res}=[2,3,4,5,6,7]$ and $N=38084+(76916/R_{SH}^{res})$. The calculation parameters for the multiscale "MS" and cubic system are:$r_{c}=1.0$, $r_s=0.3$, $r_{cell}=r_{c}+r_s$,$V=1157.625\sigma^3$, $V_{HR}=140.608\sigma^3$, $V_{LR}=V-V_{HR}$ and $P=64$. The calculation parameters for the inhomogeneous "IH" and noncubic system (b) are:$r_{c}=0.9$, $r_{cell}=r_{c}$, $R_{SH}^{res}=6$, $N=20828+(124964/R_{SH}^{res})$, $V=8192\sigma^3$, $V_{HR}=4096\sigma^3$ and $V_{LR}=V-V_{HR}$. All in reduced units with $\sigma=\epsilon=m=1$.}
\label{fig:DRD_sDD_theoryVars}
\end{figure}

\begin{figure}[!hbt]
\centering
\includegraphics[clip,width=0.97\columnwidth,keepaspectratio]{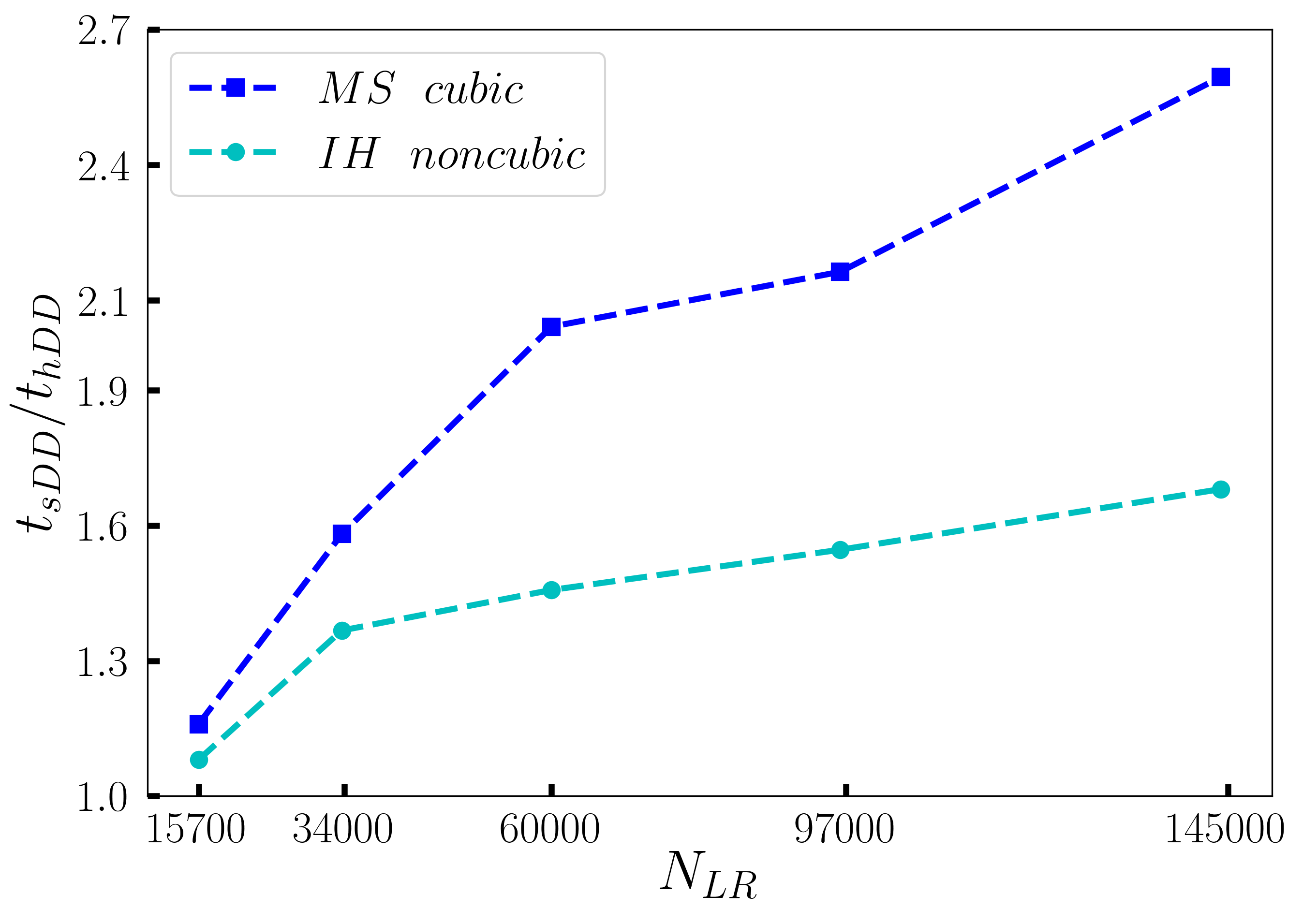}
\caption{Theoretical calculation of the computational time for the spatial domain-decomposition algorithm (see Equation~\ref{eq:sDD_scaling}) over the HeSpaDDA one (see Equation~\ref{eq:DRD_scaling}), as a function of variables (b) volume $V$ and hence $N$ in order to maintain a constant density with the specific simulation parameters: $N^{LR}=N^{0}_{LR}(V_T/V^0_T)$ with $V^0_T=1157$, $N^{0}_{LR}=38084$ and $N_{HR}=76916$. The calculation parameters for the multiscale "MS" and cubic system are:$r_{c}=1.0$, $r_s=0.3$, $r_{cell}=r_{c}+r_s$, $V_{HR}=5.2^3$, $V_{LR}=V-V_{HR}$ and $P=160$. The calculation parameters for the inhomogeneous "IH" and noncubic system (b) are:$r_{c}=0.9$, $r_{cell}=r_{c}$, $N^{LR}=N^{0}_{LR}(V_T/V^0_T)$ with $V^0_T=32\times 16\times 16$, $N^{0}_{LR}=20828$ and $N_{HR}=124964$, $V_{HR}=V/2$ and $V_{LR}=V-V_{HR}$. All in reduced units with $\sigma=\epsilon=m=1$.}
\label{fig:DRD_sDD_theoryVars2}
\end{figure}

Also from the modeling perspective, the ratio $t_{sDD}/t_{hDD}$ varies depending on the tackled system, for the cubic system a HeSpaDDA reaches a factor $\approx$ 1.5 (Figure~\ref{fig:DRD_sDD_theory}a) while for the noncubic system this factor is $\approx$ 1.3 (Figure~\ref{fig:DRD_sDD_theory}b). These results suggest that decomposing the total number of processors in a balanced way as a function of the different system resolutions and asymmetric simulation box axes has a higher limitation since the processors triplets shall be integer \emph{i.e.} ${P_x,P_y,P_z} \in $ Integers. In other words for cubic systems the resolutions are scaled equally in every simulation box axis while for noncubic ones this is tackled for each axis and each resolution.   
In order to further understand the scaling of multiple resolution simulations and the proposed modeling for the HeSpaDDA algorithm, we also show how the latter scales with respect to the sDD as a function of $R_{SH}^{res}$ (Figure~\ref{fig:DRD_sDD_theoryVars}) and $N$ (Figure~\ref{fig:DRD_sDD_theoryVars2}). Note that here direct calculations have been performed which are not taking into account any platform dependent communication patterns.
%
In Figure~\ref{fig:DRD_sDD_theoryVars} we explore how the relative speedup $t_{sDD}/t_{hDD}$ performs as a function of a variable resolution ratio $R_{SH}^{res}$ which in this illustrative scenario varies from 2:1 to 7:1.

According to this modeling results we show how the relative speedup increases monotonically with $R_{SH}^{res}$ for both systems. Although the increment of the noncubic system is significantly lower than the one given for the cubic system, as explained earlier this is due to the sensitivity on the different resolutions in each axis of the simulation box. moreover for the cubic case 3 axes are dual-resolution, while for the noncubic one, only one axis is prescribed to be dual-resolution (see Figure~\ref{fig: GeomConsDRD}) of the Appendix~\ref{appA}). We also observe a monotonic increment in relative speedup as long as the low-resolution region grows and the high-resolution region remains constant for both systems, which is plotted as a function of total number of particles in Figure~\ref{fig:DRD_sDD_theoryVars2}. It is worth noting that the scaling for the noncubic system as a function of the increment of particles in the low resolution region goes qualitatively in-line with the cubic system but presenting less speedup. The reason of such behavior is that the noncubic system in terms of processor allocation grows only in one of the simulation box axes (see also Figure~\ref{fig: LJ6to1Basic}).

A clear limitation of the scaling laws presented in this Section is that the underlying communication pattern of the computing platform is unknown and thus no estimations of the platform related load imbalances can be predictively taken into account. Unless the employed MD package includes a very fine tuned dynamic load balancing algorithm. Despite this fact the information given by the scaling Equation~\ref{eq:DRD_scaling} serve as qualitative lower boundary for estimating the duration of inhomogeneous simulation as a function of the modeled variables.

\subsection{Algorithm description}
\label{algoDes}
\subsubsection{Processor allocation}
\label{ProcAllo}
The proper allocation of processors in heterogeneous molecular simulations is crucial for the intrinsic computational scaling and performance of the production run. In other words, the computational scaling of heterogeneous simulation at this stage is not limited by the MD package used. However it is constrained to the initial domain decomposition and hence the correspondence of the number of processors to different resolution regions of the initial given configuration. One of the aspects to consider for spatially decomposed systems is the dimensional sensitivity which is the parallelepiped characteristics of the simulation setups \emph{i.e.} cubic or noncubic. A second aspect to consider is the \emph{spatial heterogeneity} ratio $R_{SH}^{res}$ (as defined in Equation~\ref{eq:R_MS}) and the third is definitely the volumes of high and low resolution regions $V_{HR}$ and $V_{LR}$, respectively. More details about the algorithm flow chart and implementation are given in the Appendix~\ref{appC}.

\subsubsection{Cell partitioning}
Once the three-dimensional processors grid has been built, cell partitioning is required (this flow chart is illustrated in the Appendix~\ref{appA}). To this end, the module for returning the inhomogeneously distributed cells along each dimension of the box is used. Within this module the precise ``load'' in terms of number of cells are allocated to each processor. For achieving the cells partitioning, the module requires the number of processors, $R_{SH}^{res}$ and the shape of the heterogeneous system per dimension. In some heterogeneous simulation setups, the same amount of processors as amount of cells could be given. Thereby the algorithm verifies such setups and resolves if the distribution of cells per processors could be treated homogeneously or not. In case of homogeneity, the strict linked-cell-list partitioning is employed.

The cells redistribution is performed by finding the integer quotient of the number of processors divided by the number of cells \emph{i.e.} $q_{HR}=\frac{C_{HR}}{P_{HR}}$ and $q_{LR}=\frac{C_{LR}}{P_{LR}}$. On the contrary to the processor allocation module from Section~\ref{ProcAllo}, the cells partitioning makes emphasis on the low-resolution region. In other words, for the less expensive region (low-resolution) the quotient of cells per processor $q_{LR} > q_{HR}$. 

Mathematically the cells in the low-resolution are weighted as a function of the resolution ratio $R_{SH}^{res}$ and the volumes $V_{LR}$ and $V_{HR}$. Subject to the numbers of processors and cells there are cases where no integer quotient is found so that HeSpaDDA rounds down the real quotient solution and redistribute the ``residual cells''. The latter type of cells are distributed using a ``pseudo-random'' mechanisms as detailed in the Appendix~\ref{appA}.

With the goal of making the algorithm available for different heterogeneous simulation techniques and MD simulation packages, the algorithm is implemented in ESPResSo++ and is also available as a stand-alone Python script~\cite{GHH}.

\section{Simulations}
\label{simDetails}
The exhibited challenges in terms of parallelization schemes for heterogeneous molecular simulations have been thoroughly studied by using two archetypical systems, namely, an ubiquitin protein solvated in water and the binary Lennard-Jones fluid. Other technical details like the MD packages versions and underneath hardware platform used are found in the Appendix~\ref{appC}. 

\subsection{Ubiquitin in aqueous solution}
\label{ubiquitin}
The system is illustrated in Figure~\ref{fig: adress app} and contained one protein molecule solvated in 38084 water molecules. The simulation box was a cube of length $\approx$10.5 nm. In terms of the cell-grid, a cubic decomposition of the system turns into a number of cells per simulation box side of 8 and hence cell-grid $(8 \times 8 \times 8)$, where each cell has a side length of 1.3 nm ($r_c+r_{s}$). The atomistic region was a sphere of radius 2 nm centered on a protein atom, such that the protein fits the sphere and thus was completely atomistic at all times. The width of the hybrid region was 1.0 nm. The benchmarking simulations presented here were performed using the simulation parameters given in a previous work~\cite{Kremer_JCP_2015-adresprot}. The coarse-grained solvent model was obtained via Iterative Boltzmann Inversion (IBI)~\cite{Soper_JCP_1996,KK_AcPoly_1998,Reith2003}. Non-bonded atomistic and coarse-grained interactions were interpolated using the Force-AdResS approach. The non-bonded cutoff was 1.0 nm, and a thermodynamic force was used to counteract the pressure difference between atomistic and coarse-grained forcefields.

\subsection{Phase separated Lennard-Jones binary fluid}
\label{1to6Section}

The second simulation setup for benchmarking was a single-scale Lennard-Jones system of a phase separated 6 to 1 $\sigma$ binary fluid. All non-bonded interactions were cut off at $0.9$ cutoff and interaction potentials shifted such that the energies are zero at the cutoff. The simulations were performed in the canonical ensemble at a temperature of $300\,K$, for which we used a Langevin thermostat with friction constant $\gamma = 0.5\,\textrm{ps}^{-1}$. The total number of particles in the system is 145792, with 124964 particles located in the high-resolution and 20828 in th low-resolution regions. The box dimensions were $32.0\times 16.0\times 16.0$ with periodic boundary conditions in each direction. Both half of the box were equilibrated separately and then together. This way we are starting out with the maximal load imbalance.

\section{Results}
\label{results}

\subsection{Ubiquitin in aqueous solution}
\label{ubiResults}
\begin{figure}[!t]
\centering
\includegraphics[clip,width=1.0\columnwidth,keepaspectratio]{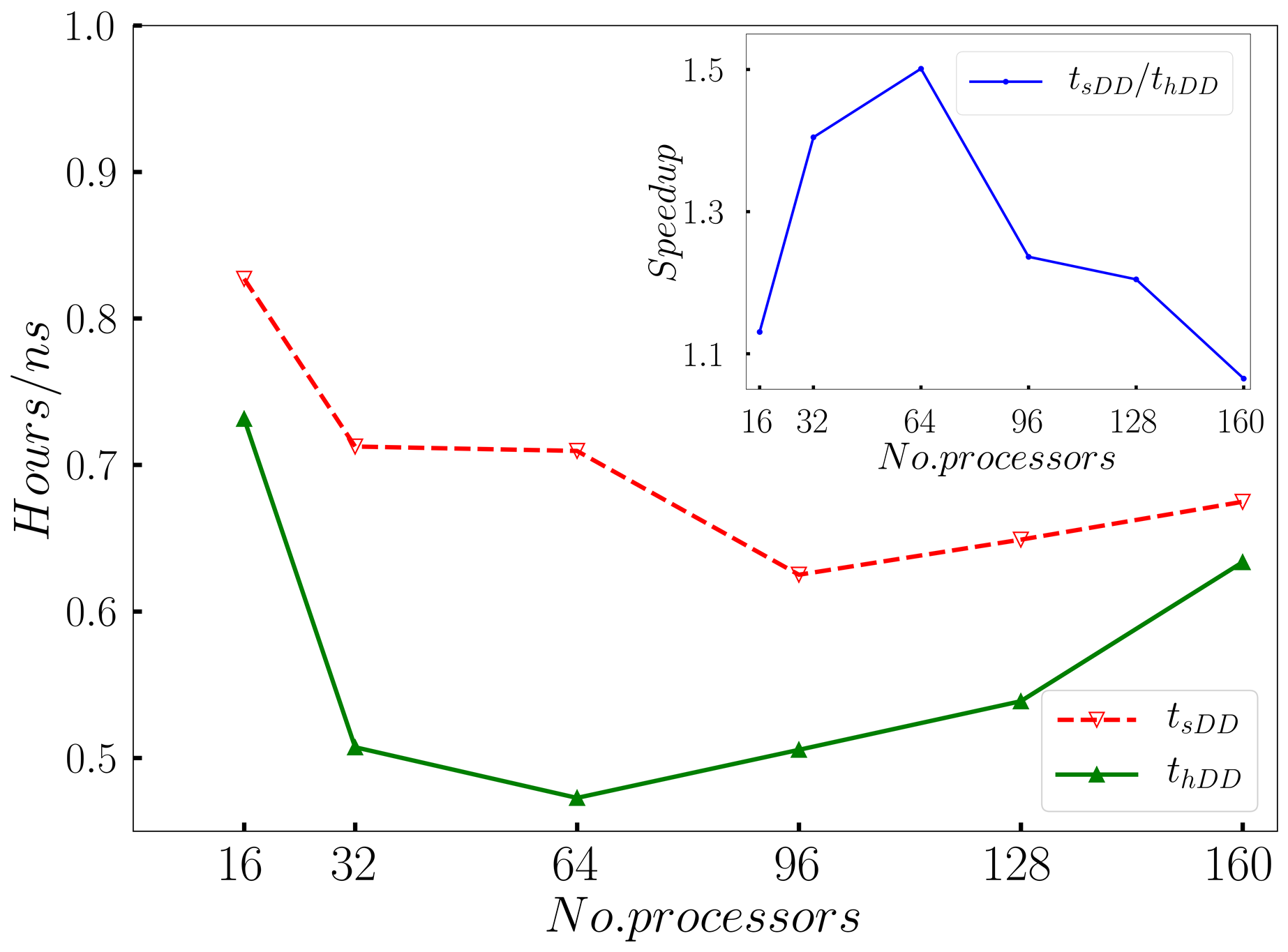}
\caption{Force-AdResS simulation of an atomistic protein and its atomistic hydration layers.The comparison between the simulation time of the spatial DD (dashed line in red) and HeSpaDDA (line in green); normalized by performance $Hours/ns$. The inset shows the ratio between the spatial DD and HeSpaDDA which corresponds to the speedup reached by the new algorithm.}
\label{fig:Ubi_results}
\end{figure}

The first system is comprised of roughly 115k atoms from which $\approx$ 88\% are coarse-grained particles build out of water molecules. In Figure~\ref{fig:Ubi_results}, the strong scaling results are shown for both algorithms sDD and HeSpaDDA in terms of performance $Hours/ns$. The latter is given by the simulation time of each value of $P=[16,32,64,96,128,160]$ required to perform a simulation for 1 ns. In all cases for the strong scaling validation shown in Figure~\ref{fig:Ubi_results}, the HeSpaDDA algorithm is much faster than sDD up to a speedup factor of 1.5. From the scaling viewpoint, HeSpaDDA reaches the best performance with 64 processors while sDD does with 96 (inset in Figure~\ref{fig:Ubi_results}).

We observe that the optimal number of processors for speeding up the first system goes along with both the shape of the box and the distribution of particles in the processors per resolution region. In other words if we tackle a cubic simulation domain the best processors-grid distribution will be a perfect cube, which in the HeSpaDDA case justifies the value of 64 ($4 \times 4 \times 4$). On the other hand sDD reaches the minimum simulation time with 96 processors which does not fulfill any system cubicity constraints nor the low and high resolution regions demarcation. After reaching the minimum for both algorithms the speedup factor among them decreases fast, and reaches $\approx$ 1.08 for 160 processors (referring to the inset of Figure \ref{fig:Ubi_results}). The reasons for the relative speedup reduction employing for example 160 processors are mainly three: 1) this number of processors cannot be decomposed as a perfect cube which increases the subdomains communication overhead; 2) the amount of particles per processor is decreased to roughly 230 which also increases the parallelization book keeping and 3) during the simulation runtime imbalances are generated due to the fact that the ubiquitin solvated in water example does not provide any dynamic load balancing algorithm. We also aimed to represent a system without dynamic load balancing and compare it to another one that includes dynamic load balancing (see Section~\ref{1to6Results}).      

As suggested in the modeling part of Section~\ref{modeling}, the scaling behavior of cubic setups with larger low-resolution regions and using the HeSpaDDA algorithm is expected to be much faster compared to the sDD one. However, the current system reflects a worst-case scenario from the domain decomposition viewpoint, because the number of cells per box lengths $(l_x,l_y,l_z)$ is 8 with 5 cells under influence of the atomistic and hybrid regions and only 3 remaining cells are left for cells partitioning in the low-resolution region. The values of number of cells are calculated by directly dividing the box lengths by $(r_c+r_s)$.

\subsection{Phase separated Lennard-Jones binary fluid}
\label{1to6Results}
The second simulation setup for benchmarking was a single-scale Lennard-Jones system of a phase separated binary fluid. The heterogeneity ratio $R_{SH}^{res}$ is 6:1 based on $\sigma$. Such system comprises $\approx$145k particles where 50\% are low resolution built out as the ratio $R_{SH}^{res}$.

\begin{figure}[!t]
\centering
\includegraphics[clip,width=1.0\columnwidth,keepaspectratio]{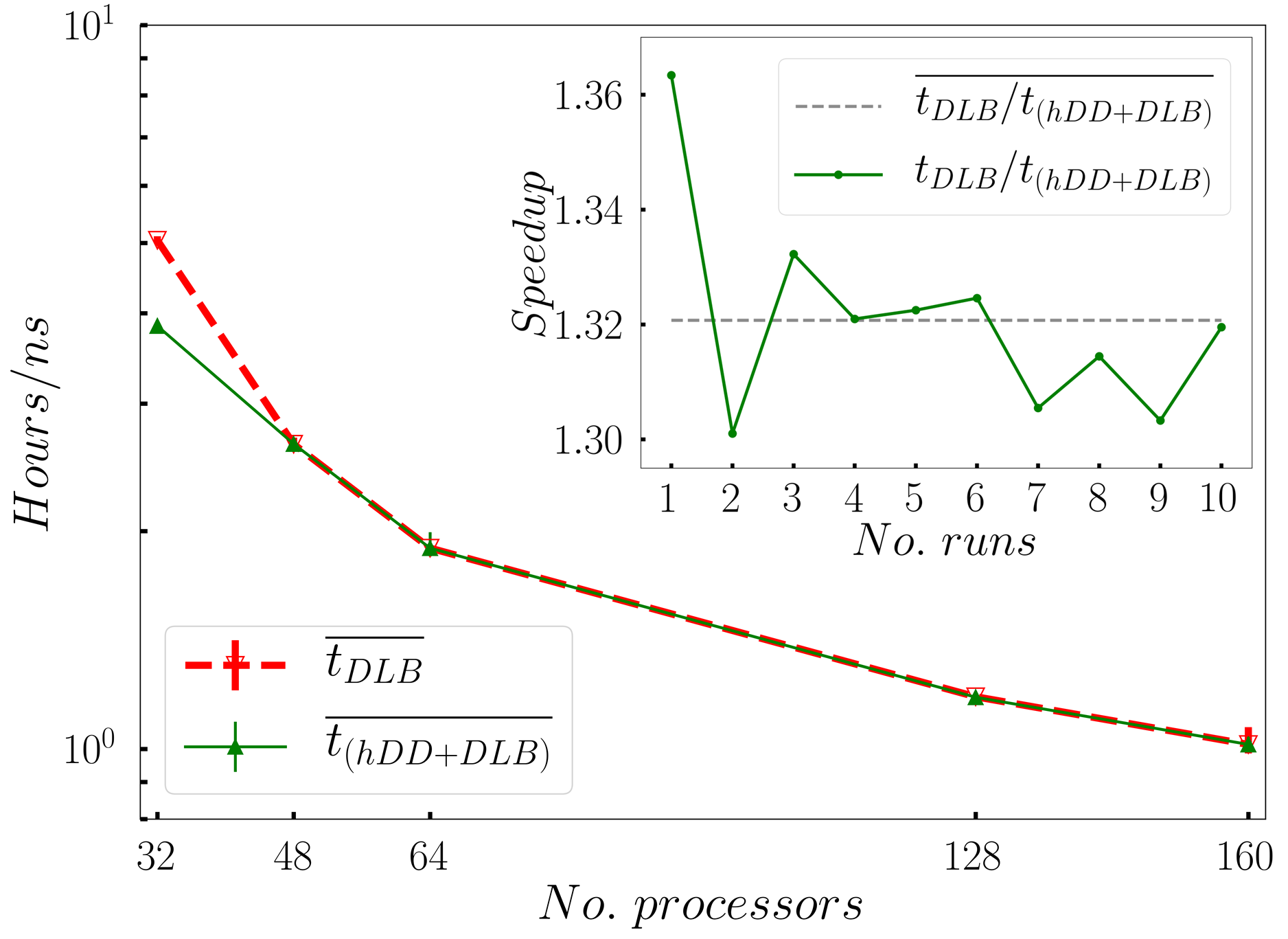}
\caption{Phase separated Lennard-Jones binary fluid.The comparison between the simulation time of GROMACS-2016 load balancing (dashed and thick line in red) and HeSpaDDA combined with GROMACS-2016 load balancing (line in green); normalized by performance $Hours/ns$. The inset shows the ratio between the GROMACS-2016 load balancing and HeSpaDDA combined with GROMACS-2016 load balancing which corresponds to the speedup reached by the new algorithm.}
\label{fig:LJ6to1Results}
\end{figure}

In Figure~\ref{fig:LJ6to1Results}, we show a direct comparison of the strong scaling simulations performed for both algorithms HeSpaDDA and GROMACS-2016~\cite{Gromacs2015} domain decomposition with notation "\emph{DLB}". Interestingly, for this system we combined HeSpaDDA to the DLB algorithm included in GROMACS-2016 with notation "\emph{hDD+DLB}" (Figure~\ref{fig:LJ6to1Results}).
The results are again shown in terms of performance $Hours/ns$ and the strong scaling range is given as $P=[32,48,64,128,160]$. These results indicate that for $P=[48,64,128,160]$ both algorithms show the same performance within the error bars given in Figure~\ref{fig:LJ6to1Results}. 

Similar results have been achieved because the initial processors triplets $(P_x,P_y,P_z)$ given by both algorithms at the beginning of the simulation gave exactly the same values. In other words, the percolation of triplets to find the total number of processors for $P=[48,64,128,160]$ is limited in those cases given the dimensions of the noncubic box. However, for $P=32$ the triplets given by the combination of HeSpaDDA and GROMACS-2016 dynamic load balancing shows an speedup in the performance of a factor $\approx$1.32 (see the inset in Figure~\ref{fig:LJ6to1Results}). Curiously the use of HeSpaDDA does not include any overhead cost, nor implementation and it works as mentioned before in a predictive manner. It is important to remark that no changes to GROMACS-2016 have been performed but changing the initial processors triplets form the command line. Hence only the processor allocation module of HeSpaDDA has been used and the tuning of cells distribution relies on the one given by GROMACS-2016.

Another interesting aspect observed in Figure~\ref{fig:LJ6to1Results} is the strong scaling of the system as a function of $P$ which is clearly superior than the previously tackled system (see Section~\ref{ubiResults}). A direct explanation for such observation comes from the fact that in the phase separated Lennard-Jones binary fluid DLB features are enabled while the previous system is configured purely with static cells and processor allocations.

\section{Discussion and Conclusion}
\label{conclusions}
While traditional molecular simulations are mostly performed with homogeneous resolution setups for all molecules in a simulation box, in heterogeneous schemes, the simulation box is typically divided in at least two regions, namely, the low-resolution and high-resolution ones. We have developed an algorithm to achieve such domain decomposition requirements by adding resolution-sensitivity to the spatial domain decomposition and combining it with an initial rearrangement of the subdomain-walls in terms of cells per processor. We come closer to the particular requirements of adaptive resolution schemes in terms of scalability and relative speedups for archetypical examples an AdResS biomolecule in solution and a binary Lennard-Jones fluid. Remarkably the tackled AdResS setup represents an archetypical system from the domain decomposition viewpoint where the low-resolution region has parsimonious dimensions. However HeSpaDDA is faster by a factor up to 1.5 than the spatial DD algorithm. A modeling framework has been also provided for HeSpaDDA in Section~\ref{modeling}. We recommend to use such predictive modeling in an \textit{a priori} way before starting to simulate any heterogeneous system like the ones described here.

Our results also show that for the binary Lennard-Jones fluid, HeSpaDDA combined to an efficient DLB algorithm can reach a speedup factor of $\approx$1.32. 

HeSpaDDA algorithm is conceived to assign a parsimonious amount of processors for the whole simulation production run. This makes an heterogeneous system as scalable as the condition $\frac{N_{HR}}{P_{HR}}=\frac{N_{LR}}{P_{LR}}$ will allow. Furthermore, we observed that HeSpaDDA speeds up the initial run of any heterogeneous simulations. In other words, multiscale simulations are faster from their beginning and ready to be tackled with further dynamic load-balancing alternatives for the production run where the underlying hardware of the HPC environment may introduce runtime imbalance, as mentioned in Section~\ref{mrDD} and shown in Section~\ref{results}.

In addition to the benchmark results, a theoretical modeling for the algorithm and the scaling law of computation time are provided. These scaling laws are favorable for exploring the upper boundaries in terms of scalability as a function of the system size or multiscale resolution ratio. Consequently mathematical details of the HeSpaDDA are presented, as well as, the algorithm flow charts and implementation. With the final goal that HeSpaDDA algorithm could be used in other multiscale techniques and/or other MD packages.

The envisioned applications for the HeSpaDDA algorithm on multiple resolution methods aim to increase scalability, and hence make larger biomolecular and advanced materials simulations more feasible. 

Besides the direct extensions of the present work to new systems of multiple spatial resolution and or single-scale inhomogeneous binary mixtures, the algorithm developed here and the referenced systems paves the way to studying more complex systems, such as evaporation, crystallization and active matter processes whenever they accomplish the slowly varying system assumption.

On top of the aspects described above, the HeSpaDDA method does not introduce any time overhead to the simulation setup nor the simulation run since it is a highly optimized algebraic function. This has been shown in Section~\ref{1to6Results} where HeSpaDDA is complementary to a dynamic load balancing technique at no additional time overhead.

\appendix

\section{HeSpaDDA algorithm description}
\label{appA}
\subsection{Processor allocation}
\begin{figure}[!b]
\centering
\includegraphics[clip,width=0.92\columnwidth,keepaspectratio]{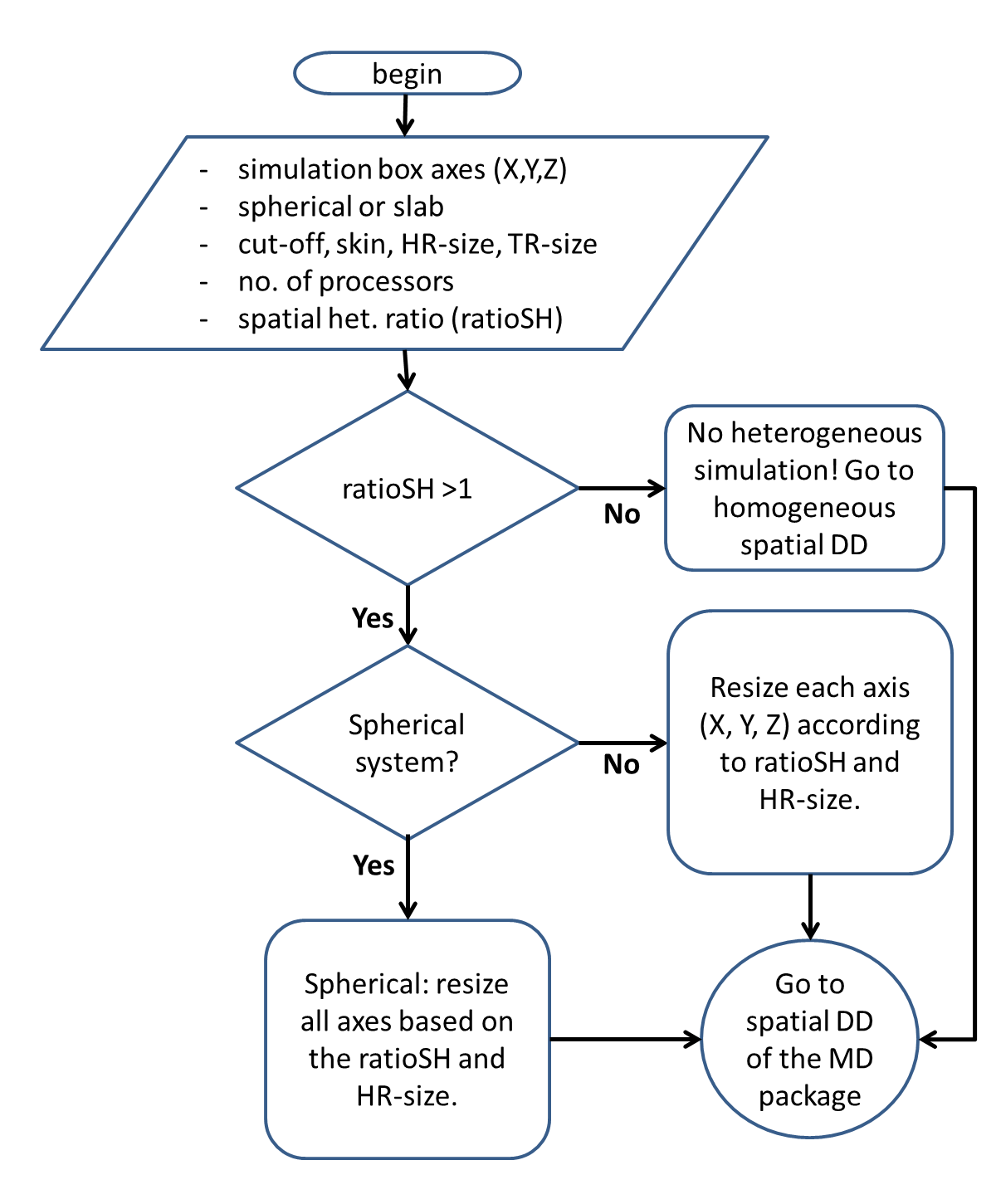}
\caption{Flow chart of the first module within HeSpaDDA algorithm devoted to the allocation of processors along the heterogeneous simulation setup.The present module provides the processors grid in terms of $P_x,P_y,P_z$.}
\label{fig: DRD_module1}
\end{figure}

We start by describing the HeSpaDDA algorithm flow chart for the processor allocation as illustrated in Figure~\ref{fig: DRD_module1}. In terms of processors, HeSpaDDA is allocating higher priority to the high-resolution regions according to Equation $P_{HR}^{hDD}=\frac{R_{SH}^{res}V_{HR}}{V+V_{HR}(R_{SH}^{res}-1)}P$ as described in the modeling Section of this article. Consequently, more processors-per-cells will be available in the box region where the high-resolution is located than in the low-resolution one. In addition, the overall allocation of processors per simulation box axis is thoroughly controlled by verifying that the total number of processors in each simulation box axis does not exceed the number of cells per box axis (\emph{i. e.} $X,Y,Z$).

\begin{figure}[!h]
\centering
\includegraphics[clip,width=1.04\columnwidth,keepaspectratio]{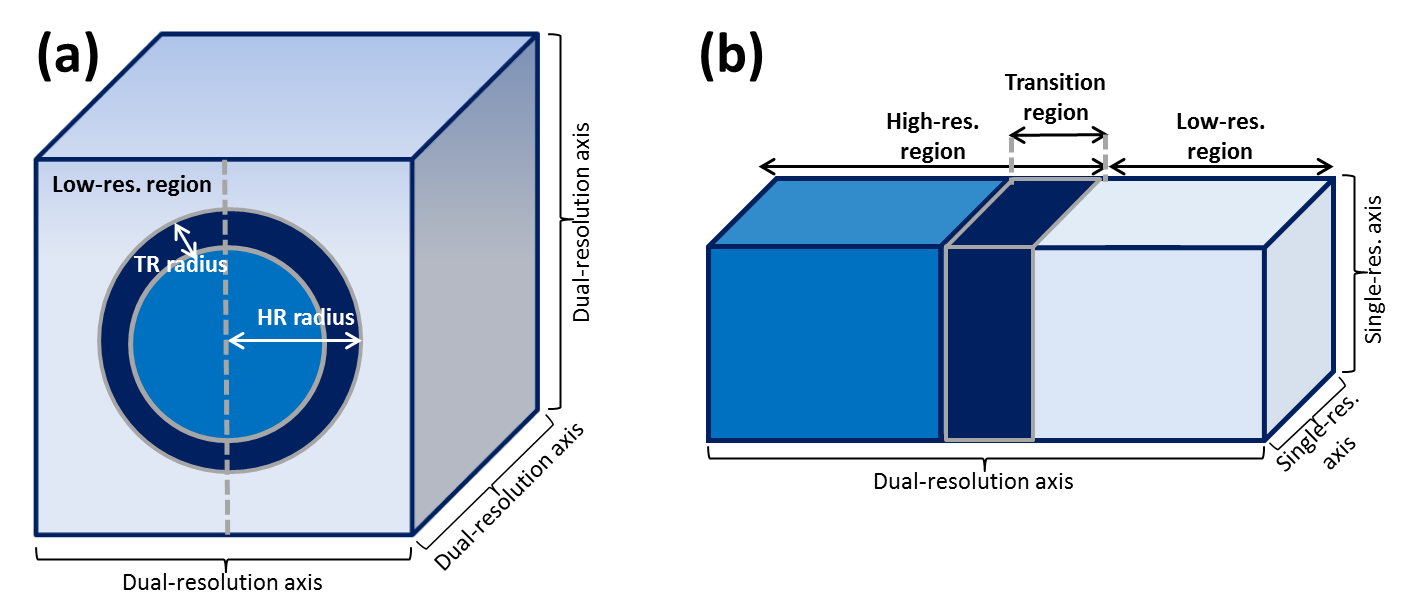}
\caption{Description of the geometrical characteristics of the heterogeneous molecular systems like (a) cubic and spherical; (b) noncubic and slab-like. The illustration also denotes which axis in a certain type of geometrical configuration will be considered as single or dual resolution.}
\label{fig: GeomConsDRD}
\end{figure}

For slab-like high-resolution configurations there is a supplementary control flow which based on the $R_{SH}^{res}$ and the number of processors per axis (\emph{i. e.} $P_x, P_y, P_z$) distinguishes where resources should be assigned according to a single(homogeneous) or dual resolutions(heterogeneous) as depicted in Figure~\ref{fig: GeomConsDRD}. Another example of a single resolution spatial DD within HeSpaDDA is having a small number of processors per axis. For example let's consider $P=8$, whereas for a cubic configuration we have $P_x=P_y=P_z=2$ and hence the processors allocation cannot be efficient under an heterogeneous DD scheme.

\subsection{Cell partitioning}
\label{CellsPart}
\begin{figure}[!t]
\centering
\includegraphics[clip,width=1.02 \columnwidth,keepaspectratio]{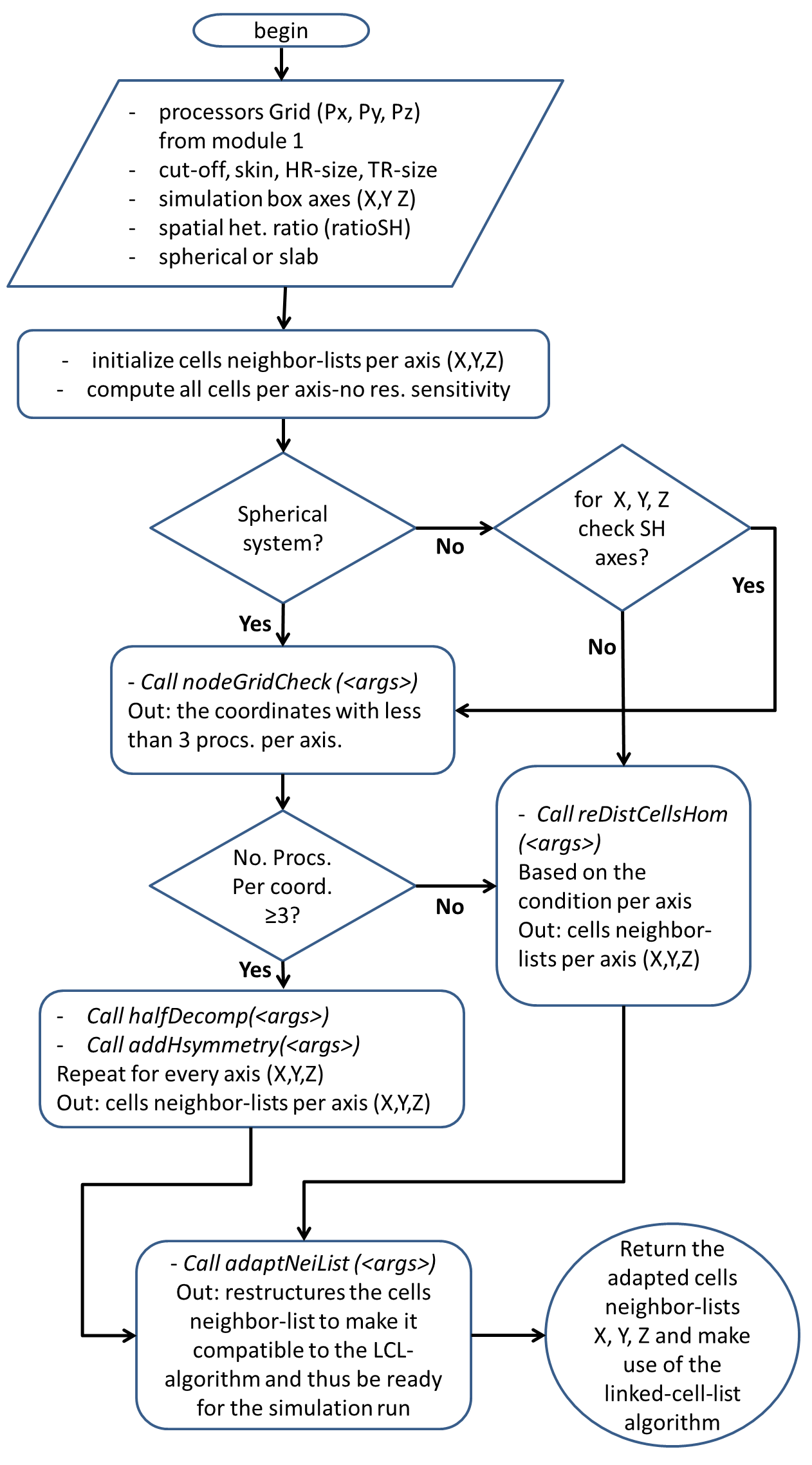}
\caption{Flow chart of the second module within HeSpaDDA algorithm devoted to distribution of cells along coordinates in each subdomain, according to the high and low resolution regions of the heterogeneous simulation setup.}
\label{fig: DRD_modul2}
\end{figure}

Once the three-dimensional processors grid has been built, cell partitioning is required (this flow chart is illustrated in Figure~\ref{fig: DRD_modul2}). To this end, the module for returning the inhomogeneously distributed cells (or cells neighbor-list) along each axis of the box is called. Within this module the precise load in terms of number of cells are allocated to each processor. For achieving the cell partitioning, the module requires the number of processors, $R_{SH}^{res}$ and the shape of the heterogeneous system per box axis. In some heterogeneous simulation setups, the same amount of processors as amount of cells could be given. Thereby the algorithm verifies such setups and resolves if the distribution of cells per processors could be treated homogeneously or not. In case of homogeneity, the strict linked-cell-list partitioning will be applied.

For heterogeneous systems where the total number of cells and processors differ, a new cells distribution scheme is presented. Such scheme assumes in forehand that the total number of cells in the system can be symmetrically divided by two for each box axis \emph{i. e. X, Y, Z}. Hence a first function is called to start the half-size decomposition (in the flowchart of Figure~\ref{fig: DRD_modul2} found as:\textit{halfDecomp}).

The advantage of the half-symmetric decomposition appears when the symmetry condition is fulfilled and thus the ``half-decomposed system'' can be mirrored. Consequently a rapid whole domain decomposition will be achieved. However, HeSpaDDA is also able to tackle asymmetric cases by means of the subsequent cells redistribution (which can be found in Figure~\ref{fig: DRD_modul2} as \textit{addHsymmetry}). The cells redistribution is performed by finding the integer quotient of the number of cells divided by the number of processors \emph{i.e.} $C_{HR}/P_{HR}$ and $C_{LR}/P_{LR}$. Contrary to the processor allocation module from Section~\ref{ProcAllo}, the cells partitioning assigns more weight in terms of cells to the low-resolution region. In other words, for the less expensive region (low-resolution) the ratio of cells per processor $C_{LR}/P_{LR}$ is mostly higher that $C_{HR}/P_{HR}$. 

Mathematically the cells in the low-resolution region are weighted as a function of the resolution ratio $R^{res}_{SH}$ (\emph{RatioSH}) and the volumes $V_{LR}$ and $V_{HR}$. Subject to the numbers of processors and cells there are cases where no integer quotient is found so that HeSpaDDA rounds down the real quotient solution and redistribute the ``residual cells''. The latter type of cells are distributed using a ``pseudo-random'' mechanism. This mechanism controls the ``residual cells'' distribution by assigning flags to the processors that already contain one ``residual cell'', so that, they will not repeatedly be assigned to the same processor. In the code~\cite{GHH} such functions are embedded in the \textit{addHsymmetry} function (see Figure~\ref{fig: DRD_modul2}). As a final step the algorithm will adapt the current cells partitioning to the neighbor-list data structure employed in a given simulation package. For example a code based on the Linked-Cell-List algorithm.

The algorithm is also sensitive to ill-conditioned heterogeneous setup and therefore messages will be displayed as a warning of missing scalability or performance.
%

In Figure~\ref{fig: GeomConsDRD}(a), the center of the simulation box correspond to the center of the high-resolution region. However, if the initial simulation setup is not placing the high-resolution region in the center of the simulation box, HeSpaDDA calculates the offset between those centers by considering periodic boundary conditions and hence the cells partitioning will be shifted by the nearest integer quotient offset/$r_{cell}$.

With the goal of making the algorithm available for different heterogeneous simulation techniques and corresponding MD simulation packages, the algorithm is implemented in ESPResSo++ and is available as an stand-alone Python script.

\section{Simulation setup}
\label{appB}
In the presented work, two systems have been tackled the multiscale protein solvated in water and the Lennard-Jones binary fluid.
\subsection{Ubiquitin in aqueous solution}
The simulations have been carried out by building the codes with Intel MPI compiler v14.0, Infiniband-Cluster and two versions of ESPResSo++: the first (v1.9.4) uses an spatial Domain Decomposition while the second is publicly available within this paper also on ESPResSo++ github. Each simulation was run for ten thousand steps and with a variable number of processors ($P=[16,32,64,96,128,160]$). The skin was optimized for each system and for $P=16$ (using the \emph{TuneSkin} function of ESPResSo++ v1.9.4). $dt$ of the protein in solution was 1$fs$. Only the time spent integrating the equations of motion was taken into account \emph{i.e.} no file I/O was considered. Note also that the values given are the average times of a total of 5 runs for each data point in the article (see Section~\ref{results}).
\subsection{Phase separated Lennard-Jones binary fluid}
For the second system, the simulations have been carried out by building the codes with Intel MPI compiler v14.0, Infiniband-Cluster and one version of GROMACS-2016  which has an embedded dynamic load balancing algorithm that has been employed. Note that for the simulations with GROMACS, there are not two versions because such package allows to chose the number of processors per simulation box axis in advance. Each simulation was run for ten thousand steps with a variable number of processors ($P=[32,64,128,160]$) and $dt$ of the  1$fs$. Only the time spent integrating the equations of motion was taken into account \emph{i.e.} no file I/O was considered. Note also that the values given are the average times of a total of 10 runs for each data point in the article (see Section~\ref{results}). We have added more runs in this second case since we tackle a simulation with the dynamic load balancing feature enabled.

\section{Implementation}
\label{appC}
The validation and benchmarks have been carried out after a new implementation on the current release (v1.9.4) of the ESPResSo++ package. This release offers an homogeneous-spatial domain decomposition scheme combined with the linked-cell-list algorithm~\cite{Halverson2013}.
The HeSpaDDA algorithm needed the following backward compatible modifications to ESPResSo++ (v1.9.4): new data structure for the subdomains cells neighbor-lists \emph{neiListX}, \emph{neiListY} and \emph{neiListZ}, an iterative function that allocates processors and one function that distributes cells according to the given resolution types per box axis X,Y,Z and \emph{ratioSH}.
We have provided a brief overview of those modifications, while further code details can be found in the ESPResSo++ github repository:~\url{https://github.com/espressopp/espressopp}.

\begin{acknowledgments}
T. Stuehn and H. V. Guzman acknowledges financial support under the project SFB-TRR146 of the Deutsche Forschungsgemeinschaft. The authors are thankful to Jan Smrek, Karsten Kreis and Aoife Fogarty for stimulating and fruitful discussions on the manuscript.
\end{acknowledgments}
\bibliographystyle{apsrev4-1}
\bibliography{pubs2}

\begin{thebibliography}{57}%
\makeatletter
\providecommand \@ifxundefined [1]{%
 \@ifx{#1\undefined}
}%
\providecommand \@ifnum [1]{%
 \ifnum #1\expandafter \@firstoftwo
 \else \expandafter \@secondoftwo
 \fi
}%
\providecommand \@ifx [1]{%
 \ifx #1\expandafter \@firstoftwo
 \else \expandafter \@secondoftwo
 \fi
}%
\providecommand \natexlab [1]{#1}%
\providecommand \enquote  [1]{``#1''}%
\providecommand \bibnamefont  [1]{#1}%
\providecommand \bibfnamefont [1]{#1}%
\providecommand \citenamefont [1]{#1}%
\providecommand \href@noop [0]{\@secondoftwo}%
\providecommand \href [0]{\begingroup \@sanitize@url \@href}%
\providecommand \@href[1]{\@@startlink{#1}\@@href}%
\providecommand \@@href[1]{\endgroup#1\@@endlink}%
\providecommand \@sanitize@url [0]{\catcode `\\12\catcode `\$12\catcode
  `\&12\catcode `\#12\catcode `\^12\catcode `\_12\catcode `\%12\relax}%
\providecommand \@@startlink[1]{}%
\providecommand \@@endlink[0]{}%
\providecommand \url  [0]{\begingroup\@sanitize@url \@url }%
\providecommand \@url [1]{\endgroup\@href {#1}{\urlprefix }}%
\providecommand \urlprefix  [0]{URL }%
\providecommand \Eprint [0]{\href }%
\providecommand \doibase [0]{http://dx.doi.org/}%
\providecommand \selectlanguage [0]{\@gobble}%
\providecommand \bibinfo  [0]{\@secondoftwo}%
\providecommand \bibfield  [0]{\@secondoftwo}%
\providecommand \translation [1]{[#1]}%
\providecommand \BibitemOpen [0]{}%
\providecommand \bibitemStop [0]{}%
\providecommand \bibitemNoStop [0]{.\EOS\space}%
\providecommand \EOS [0]{\spacefactor3000\relax}%
\providecommand \BibitemShut  [1]{\csname bibitem#1\endcsname}%
\let\auto@bib@innerbib\@empty
\bibitem [{\citenamefont {Kremer}\ and\ \citenamefont
  {M{\"u}ller-Plathe}(2002)}]{KK-FMP2002}%
  \BibitemOpen
  \bibfield  {author} {\bibinfo {author} {\bibfnamefont {K.}~\bibnamefont
  {Kremer}}\ and\ \bibinfo {author} {\bibfnamefont {F.}~\bibnamefont
  {M{\"u}ller-Plathe}},\ }\href@noop {} {\bibfield  {journal} {\bibinfo
  {journal} {Molecular Simulation}\ }\textbf {\bibinfo {volume} {28}},\
  \bibinfo {pages} {729} (\bibinfo {year} {2002})}\BibitemShut {NoStop}%
\bibitem [{\citenamefont {Attig}\ \emph {et~al.}(2004)\citenamefont {Attig},
  \citenamefont {Binder}, \citenamefont {Grubm{\"u}ller},\ and\ \citenamefont
  {Kremer}}]{attig04}%
  \BibitemOpen
  \bibinfo {editor} {\bibfnamefont {N.}~\bibnamefont {Attig}}, \bibinfo
  {editor} {\bibfnamefont {K.}~\bibnamefont {Binder}}, \bibinfo {editor}
  {\bibfnamefont {H.}~\bibnamefont {Grubm{\"u}ller}}, \ and\ \bibinfo {editor}
  {\bibfnamefont {K.}~\bibnamefont {Kremer}},\ eds.,\ \href@noop {} {\emph
  {\bibinfo {title} {{Computational Soft Matter: From Synthetic Polymers to
  Proteins, NIC Lecture Notes}}}},\ Vol.~\bibinfo {volume} {23}\ (\bibinfo
  {publisher} {Forschungszentrum J{\"u}lich},\ \bibinfo {year}
  {2004})\BibitemShut {NoStop}%
\bibitem [{\citenamefont {Voth}(2008)}]{voth08}%
  \BibitemOpen
  \bibinfo {editor} {\bibfnamefont {G.~A.}\ \bibnamefont {Voth}},\ ed.,\
  \href@noop {} {\emph {\bibinfo {title} {{Coarse-Graining of Condensed Phase
  and Biomolecular Systems}}}}\ (\bibinfo  {publisher} {CRC Press, Boca Raton,
  Florida},\ \bibinfo {year} {2008})\BibitemShut {NoStop}%
\bibitem [{\citenamefont {Holm}\ and\ \citenamefont {Kremer}(2009)}]{holm0509}%
  \BibitemOpen
  \bibinfo {editor} {\bibfnamefont {C.}~\bibnamefont {Holm}}\ and\ \bibinfo
  {editor} {\bibfnamefont {K.}~\bibnamefont {Kremer}},\ eds.,\ \href@noop {}
  {\emph {\bibinfo {title} {{Advanced Computer Simulation Approaches for Soft
  Matter Science I-III, Series: Advances in Polymer Science}}}},\ Vol.\
  \bibinfo {volume} {173, 185, 221}\ (\bibinfo  {publisher} {Springer,
  Berlin},\ \bibinfo {year} {2005-2009})\BibitemShut {NoStop}%
\bibitem [{\citenamefont {Peter}\ and\ \citenamefont
  {Kremer}(2010)}]{peter_faraday_2010}%
  \BibitemOpen
  \bibfield  {author} {\bibinfo {author} {\bibfnamefont {C.}~\bibnamefont
  {Peter}}\ and\ \bibinfo {author} {\bibfnamefont {K.}~\bibnamefont {Kremer}},\
  }\href@noop {} {\bibfield  {journal} {\bibinfo  {journal} {Faraday Discuss.}\
  }\textbf {\bibinfo {volume} {144}},\ \bibinfo {pages} {9} (\bibinfo {year}
  {2010})}\BibitemShut {NoStop}%
\bibitem [{\citenamefont {Halverson}\ \emph {et~al.}(2013)\citenamefont
  {Halverson}, \citenamefont {Brandes}, \citenamefont {Lenz}, \citenamefont
  {Arnold}, \citenamefont {Bevc}, \citenamefont {Starchenko}, \citenamefont
  {Kremer}, \citenamefont {Stuehn},\ and\ \citenamefont
  {Reith}}]{Halverson2013}%
  \BibitemOpen
  \bibfield  {author} {\bibinfo {author} {\bibfnamefont {J.~D.}\ \bibnamefont
  {Halverson}}, \bibinfo {author} {\bibfnamefont {T.}~\bibnamefont {Brandes}},
  \bibinfo {author} {\bibfnamefont {O.}~\bibnamefont {Lenz}}, \bibinfo {author}
  {\bibfnamefont {A.}~\bibnamefont {Arnold}}, \bibinfo {author} {\bibfnamefont
  {S.}~\bibnamefont {Bevc}}, \bibinfo {author} {\bibfnamefont {V.}~\bibnamefont
  {Starchenko}}, \bibinfo {author} {\bibfnamefont {K.}~\bibnamefont {Kremer}},
  \bibinfo {author} {\bibfnamefont {T.}~\bibnamefont {Stuehn}}, \ and\ \bibinfo
  {author} {\bibfnamefont {D.}~\bibnamefont {Reith}},\ }\href {\doibase
  10.1016/j.cpc.2012.12.004} {\bibfield  {journal} {\bibinfo  {journal}
  {Comput. Phys. Commun.}\ }\textbf {\bibinfo {volume} {184}},\ \bibinfo
  {pages} {1129} (\bibinfo {year} {2013})}\BibitemShut {NoStop}%
\bibitem [{\citenamefont {Liu}\ \emph {et~al.}(2015)\citenamefont {Liu},
  \citenamefont {Grest}, \citenamefont {Marchetti}, \citenamefont {Grason},
  \citenamefont {Robbins}, \citenamefont {Fredrickson}, \citenamefont
  {Rubinstein},\ and\ \citenamefont {{Olvera de la Cruz}}}]{OlveraSM2k15}%
  \BibitemOpen
  \bibfield  {author} {\bibinfo {author} {\bibfnamefont {A.~J.}\ \bibnamefont
  {Liu}}, \bibinfo {author} {\bibfnamefont {G.~S.}\ \bibnamefont {Grest}},
  \bibinfo {author} {\bibfnamefont {M.~C.}\ \bibnamefont {Marchetti}}, \bibinfo
  {author} {\bibfnamefont {G.~M.}\ \bibnamefont {Grason}}, \bibinfo {author}
  {\bibfnamefont {M.~O.}\ \bibnamefont {Robbins}}, \bibinfo {author}
  {\bibfnamefont {G.~H.}\ \bibnamefont {Fredrickson}}, \bibinfo {author}
  {\bibfnamefont {M.}~\bibnamefont {Rubinstein}}, \ and\ \bibinfo {author}
  {\bibfnamefont {M.}~\bibnamefont {{Olvera de la Cruz}}},\ }\href {\doibase
  10.1039/C4SM02344G} {\bibfield  {journal} {\bibinfo  {journal} {Soft Matter}\
  }\textbf {\bibinfo {volume} {11}},\ \bibinfo {pages} {2326} (\bibinfo {year}
  {2015})}\BibitemShut {NoStop}%
\bibitem [{\citenamefont {Rapaport}(1991)}]{RapaportCPC1991}%
  \BibitemOpen
  \bibfield  {author} {\bibinfo {author} {\bibfnamefont {D.}~\bibnamefont
  {Rapaport}},\ }\href {\doibase 10.1016/0010-4655(91)90095-3} {\bibfield
  {journal} {\bibinfo  {journal} {Computer Physics Communications}\ }\textbf
  {\bibinfo {volume} {62}},\ \bibinfo {pages} {198} (\bibinfo {year}
  {1991})}\BibitemShut {NoStop}%
\bibitem [{\citenamefont {Shaw}\ \emph {et~al.}(2009)\citenamefont {Shaw},
  \citenamefont {Dror}, \citenamefont {Salmon}, \citenamefont {Grossman},
  \citenamefont {Mackenzie}, \citenamefont {Bank}, \citenamefont {Young},
  \citenamefont {Deneroff}, \citenamefont {Batson}, \citenamefont {Bowers},
  \citenamefont {Chow}, \citenamefont {Eastwood}, \citenamefont {Ierardi},
  \citenamefont {Klepeis}, \citenamefont {Kuskin}, \citenamefont {Larson},
  \citenamefont {Lindorff-Larsen}, \citenamefont {Maragakis}, \citenamefont
  {Moraes}, \citenamefont {Piana}, \citenamefont {Shan},\ and\ \citenamefont
  {Towles}}]{ShawPCHPCNSA2009}%
  \BibitemOpen
  \bibfield  {author} {\bibinfo {author} {\bibfnamefont {D.~E.}\ \bibnamefont
  {Shaw}}, \bibinfo {author} {\bibfnamefont {R.~O.}\ \bibnamefont {Dror}},
  \bibinfo {author} {\bibfnamefont {J.~K.}\ \bibnamefont {Salmon}}, \bibinfo
  {author} {\bibfnamefont {J.~P.}\ \bibnamefont {Grossman}}, \bibinfo {author}
  {\bibfnamefont {K.~M.}\ \bibnamefont {Mackenzie}}, \bibinfo {author}
  {\bibfnamefont {J.~A.}\ \bibnamefont {Bank}}, \bibinfo {author}
  {\bibfnamefont {C.}~\bibnamefont {Young}}, \bibinfo {author} {\bibfnamefont
  {M.~M.}\ \bibnamefont {Deneroff}}, \bibinfo {author} {\bibfnamefont
  {B.}~\bibnamefont {Batson}}, \bibinfo {author} {\bibfnamefont {K.~J.}\
  \bibnamefont {Bowers}}, \bibinfo {author} {\bibfnamefont {E.}~\bibnamefont
  {Chow}}, \bibinfo {author} {\bibfnamefont {M.~P.}\ \bibnamefont {Eastwood}},
  \bibinfo {author} {\bibfnamefont {D.~J.}\ \bibnamefont {Ierardi}}, \bibinfo
  {author} {\bibfnamefont {J.~L.}\ \bibnamefont {Klepeis}}, \bibinfo {author}
  {\bibfnamefont {J.~S.}\ \bibnamefont {Kuskin}}, \bibinfo {author}
  {\bibfnamefont {R.~H.}\ \bibnamefont {Larson}}, \bibinfo {author}
  {\bibfnamefont {K.}~\bibnamefont {Lindorff-Larsen}}, \bibinfo {author}
  {\bibfnamefont {P.}~\bibnamefont {Maragakis}}, \bibinfo {author}
  {\bibfnamefont {M.~A.}\ \bibnamefont {Moraes}}, \bibinfo {author}
  {\bibfnamefont {S.}~\bibnamefont {Piana}}, \bibinfo {author} {\bibfnamefont
  {Y.}~\bibnamefont {Shan}}, \ and\ \bibinfo {author} {\bibfnamefont
  {B.}~\bibnamefont {Towles}},\ }in\ \href {\doibase 10.1145/1654059.1654126}
  {\emph {\bibinfo {booktitle} {{Proceedings of the Conference on High
  Performance Computing Networking, Storage and Analysis}}}}\ (\bibinfo {year}
  {2009})\ pp.\ \bibinfo {pages} {1--11}\BibitemShut {NoStop}%
\bibitem [{\citenamefont {Li}\ \emph {et~al.}(2013)\citenamefont {Li},
  \citenamefont {Abberton}, \citenamefont {Kroeger},\ and\ \citenamefont
  {Liu}}]{KroegerP2013}%
  \BibitemOpen
  \bibfield  {author} {\bibinfo {author} {\bibfnamefont {Y.}~\bibnamefont
  {Li}}, \bibinfo {author} {\bibfnamefont {B.~C.}\ \bibnamefont {Abberton}},
  \bibinfo {author} {\bibfnamefont {M.}~\bibnamefont {Kroeger}}, \ and\
  \bibinfo {author} {\bibfnamefont {W.~K.}\ \bibnamefont {Liu}},\ }\href@noop
  {} {\bibfield  {journal} {\bibinfo  {journal} {Polymers}\ }\textbf {\bibinfo
  {volume} {5}} (\bibinfo {year} {2013})}\BibitemShut {NoStop}%
\bibitem [{\citenamefont {Rapaport}(2014)}]{RapaportJPCM2014}%
  \BibitemOpen
  \bibfield  {author} {\bibinfo {author} {\bibfnamefont {D.~C.}\ \bibnamefont
  {Rapaport}},\ }\href {\doibase 10.1088/0953-8984/26/50/503104} {\bibfield
  {journal} {\bibinfo  {journal} {J. Phys.: Condens. Matter 26 (2014) 503104}\
  } (\bibinfo {year} {2014}),\ 10.1088/0953-8984/26/50/503104},\ \Eprint
  {http://arxiv.org/abs/1409.5958v2} {arXiv:1409.5958v2 [cond-mat.soft]}
  \BibitemShut {NoStop}%
\bibitem [{\citenamefont {Piana}\ \emph {et~al.}(2014)\citenamefont {Piana},
  \citenamefont {Klepeis},\ and\ \citenamefont {Shaw}}]{ShawCOSB2014}%
  \BibitemOpen
  \bibfield  {author} {\bibinfo {author} {\bibfnamefont {S.}~\bibnamefont
  {Piana}}, \bibinfo {author} {\bibfnamefont {J.~L.}\ \bibnamefont {Klepeis}},
  \ and\ \bibinfo {author} {\bibfnamefont {D.~E.}\ \bibnamefont {Shaw}},\
  }\href {\doibase 10.1016/j.sbi.2013.12.006} {\bibfield  {journal} {\bibinfo
  {journal} {Current Opinion in Structural Biology}\ }\textbf {\bibinfo
  {volume} {24}},\ \bibinfo {pages} {98} (\bibinfo {year} {2014})},\ \bibinfo
  {note} {folding and binding / Nucleic acids and their protein
  complexes}\BibitemShut {NoStop}%
\bibitem [{\citenamefont {Field}(2015)}]{FieldABB2015}%
  \BibitemOpen
  \bibfield  {author} {\bibinfo {author} {\bibfnamefont {M.~J.}\ \bibnamefont
  {Field}},\ }\href {\doibase 10.1016/j.abb.2015.03.005} {\bibfield  {journal}
  {\bibinfo  {journal} {Archives of Biochemistry and Biophysics}\ }\textbf
  {\bibinfo {volume} {582}},\ \bibinfo {pages} {3} (\bibinfo {year} {2015})},\
  \bibinfo {note} {special issue in computational modeling on biological
  systems}\BibitemShut {NoStop}%
\bibitem [{\citenamefont {Gooneie}\ \emph {et~al.}(2017)\citenamefont
  {Gooneie}, \citenamefont {Schuschnigg},\ and\ \citenamefont
  {Holzer}}]{HolzerP2017}%
  \BibitemOpen
  \bibfield  {author} {\bibinfo {author} {\bibfnamefont {A.}~\bibnamefont
  {Gooneie}}, \bibinfo {author} {\bibfnamefont {S.}~\bibnamefont
  {Schuschnigg}}, \ and\ \bibinfo {author} {\bibfnamefont {C.}~\bibnamefont
  {Holzer}},\ }\href@noop {} {\bibfield  {journal} {\bibinfo  {journal}
  {Polymers}\ }\textbf {\bibinfo {volume} {9}} (\bibinfo {year}
  {2017})}\BibitemShut {NoStop}%
\bibitem [{\citenamefont {Praprotnik}\ \emph
  {et~al.}(2008{\natexlab{a}})\citenamefont {Praprotnik}, \citenamefont
  {Site},\ and\ \citenamefont {Kremer}}]{praprotnik_multiscale_2008}%
  \BibitemOpen
  \bibfield  {author} {\bibinfo {author} {\bibfnamefont {M.}~\bibnamefont
  {Praprotnik}}, \bibinfo {author} {\bibfnamefont {L.~D.}\ \bibnamefont
  {Site}}, \ and\ \bibinfo {author} {\bibfnamefont {K.}~\bibnamefont
  {Kremer}},\ }\href {\doibase 10.1146/annurev.physchem.59.032607.093707}
  {\bibfield  {journal} {\bibinfo  {journal} {Annual Review of Physical
  Chemistry}\ }\textbf {\bibinfo {volume} {59}},\ \bibinfo {pages} {545}
  (\bibinfo {year} {2008}{\natexlab{a}})}\BibitemShut {NoStop}%
\bibitem [{\citenamefont {Delgado-Buscalioni}\ \emph
  {et~al.}(2008)\citenamefont {Delgado-Buscalioni}, \citenamefont {Kremer},\
  and\ \citenamefont {Praprotnik}}]{delgado08a}%
  \BibitemOpen
  \bibfield  {author} {\bibinfo {author} {\bibfnamefont {R.}~\bibnamefont
  {Delgado-Buscalioni}}, \bibinfo {author} {\bibfnamefont {K.}~\bibnamefont
  {Kremer}}, \ and\ \bibinfo {author} {\bibfnamefont {M.}~\bibnamefont
  {Praprotnik}},\ }\href@noop {} {\bibfield  {journal} {\bibinfo  {journal} {J.
  Chem. Phys.}\ }\textbf {\bibinfo {volume} {128}},\ \bibinfo {pages} {114110}
  (\bibinfo {year} {2008})}\BibitemShut {NoStop}%
\bibitem [{\citenamefont {{Delgado-Buscalioni}}\ \emph
  {et~al.}(2009)\citenamefont {{Delgado-Buscalioni}}, \citenamefont {Kremer},\
  and\ \citenamefont {Praprotnik}}]{delgado-buscalioni_coupling_2009}%
  \BibitemOpen
  \bibfield  {author} {\bibinfo {author} {\bibfnamefont {R.}~\bibnamefont
  {{Delgado-Buscalioni}}}, \bibinfo {author} {\bibfnamefont {K.}~\bibnamefont
  {Kremer}}, \ and\ \bibinfo {author} {\bibfnamefont {M.}~\bibnamefont
  {Praprotnik}},\ }\href {\doibase 10.1063/1.3272265} {\bibfield  {journal}
  {\bibinfo  {journal} {The Journal of Chemical Physics}\ }\textbf {\bibinfo
  {volume} {131}},\ \bibinfo {pages} {244107} (\bibinfo {year}
  {2009})}\BibitemShut {NoStop}%
\bibitem [{\citenamefont {Potestio}\ \emph
  {et~al.}(2013{\natexlab{a}})\citenamefont {Potestio}, \citenamefont
  {Fritsch}, \citenamefont {Espa{\~n}ol}, \citenamefont {Delgado-Buscalioni},
  \citenamefont {Kremer}, \citenamefont {Everaers},\ and\ \citenamefont
  {Donadio}}]{Donadio_PRL_2013-hadres_molliq}%
  \BibitemOpen
  \bibfield  {author} {\bibinfo {author} {\bibfnamefont {R.}~\bibnamefont
  {Potestio}}, \bibinfo {author} {\bibfnamefont {S.}~\bibnamefont {Fritsch}},
  \bibinfo {author} {\bibfnamefont {P.}~\bibnamefont {Espa{\~n}ol}}, \bibinfo
  {author} {\bibfnamefont {R.}~\bibnamefont {Delgado-Buscalioni}}, \bibinfo
  {author} {\bibfnamefont {K.}~\bibnamefont {Kremer}}, \bibinfo {author}
  {\bibfnamefont {R.}~\bibnamefont {Everaers}}, \ and\ \bibinfo {author}
  {\bibfnamefont {D.}~\bibnamefont {Donadio}},\ }\href {\doibase
  10.1103/PhysRevLett.110.108301} {\bibfield  {journal} {\bibinfo  {journal}
  {Phys. Rev. Lett.}\ }\textbf {\bibinfo {volume} {110}},\ \bibinfo {pages}
  {108301} (\bibinfo {year} {2013}{\natexlab{a}})}\BibitemShut {NoStop}%
\bibitem [{\citenamefont {Smrek}\ and\ \citenamefont
  {Kremer}(2017)}]{SmrekPRL2017}%
  \BibitemOpen
  \bibfield  {author} {\bibinfo {author} {\bibfnamefont {J.}~\bibnamefont
  {Smrek}}\ and\ \bibinfo {author} {\bibfnamefont {K.}~\bibnamefont {Kremer}},\
  }\href {\doibase 10.1103/PhysRevLett.118.098002} {\bibfield  {journal}
  {\bibinfo  {journal} {Phys. Rev. Lett. 118, 098002 (2017)}\ } (\bibinfo
  {year} {2017}),\ 10.1103/PhysRevLett.118.098002},\ \Eprint
  {http://arxiv.org/abs/1701.07362v1} {arXiv:1701.07362v1 [cond-mat.soft]}
  \BibitemShut {NoStop}%
\bibitem [{\citenamefont {Radu}\ and\ \citenamefont
  {Kremer}(2017)}]{RaduPRL2017}%
  \BibitemOpen
  \bibfield  {author} {\bibinfo {author} {\bibfnamefont {M.}~\bibnamefont
  {Radu}}\ and\ \bibinfo {author} {\bibfnamefont {K.}~\bibnamefont {Kremer}},\
  }\href {\doibase 10.1103/PhysRevLett.118.055702} {\bibfield  {journal}
  {\bibinfo  {journal} {Phys. Rev. Lett.}\ }\textbf {\bibinfo {volume} {118}},\
  \bibinfo {pages} {055702} (\bibinfo {year} {2017})}\BibitemShut {NoStop}%
\bibitem [{\citenamefont {Bereau}\ and\ \citenamefont
  {Deserno}(2015)}]{BereauM2015}%
  \BibitemOpen
  \bibfield  {author} {\bibinfo {author} {\bibfnamefont {T.}~\bibnamefont
  {Bereau}}\ and\ \bibinfo {author} {\bibfnamefont {M.}~\bibnamefont
  {Deserno}},\ }\href {\doibase 10.1007/s00232-014-9738-9} {\bibfield
  {journal} {\bibinfo  {journal} {The Journal of Membrane Biology}\ }\textbf
  {\bibinfo {volume} {248}},\ \bibinfo {pages} {395} (\bibinfo {year}
  {2015})}\BibitemShut {NoStop}%
\bibitem [{\citenamefont {Pluhackova}\ and\ \citenamefont
  {B{\"o}ckmann}(2015)}]{PluhackovaBM2015}%
  \BibitemOpen
  \bibfield  {author} {\bibinfo {author} {\bibfnamefont {K.}~\bibnamefont
  {Pluhackova}}\ and\ \bibinfo {author} {\bibfnamefont {R.~A.}\ \bibnamefont
  {B{\"o}ckmann}},\ }\href {http://stacks.iop.org/0953-8984/27/i=32/a=323103}
  {\bibfield  {journal} {\bibinfo  {journal} {Journal of Physics: Condensed
  Matter}\ }\textbf {\bibinfo {volume} {27}},\ \bibinfo {pages} {323103}
  (\bibinfo {year} {2015})}\BibitemShut {NoStop}%
\bibitem [{\citenamefont {Plimpton}(1995)}]{lammps95}%
  \BibitemOpen
  \bibfield  {author} {\bibinfo {author} {\bibfnamefont {S.}~\bibnamefont
  {Plimpton}},\ }\href@noop {} {\bibfield  {journal} {\bibinfo  {journal} {J.
  Comp. Phys.}\ }\textbf {\bibinfo {volume} {117}} (\bibinfo {year}
  {1995})}\BibitemShut {NoStop}%
\bibitem [{\citenamefont {Shaw}(2005)}]{de_shaw}%
  \BibitemOpen
  \bibfield  {author} {\bibinfo {author} {\bibfnamefont {D.}~\bibnamefont
  {Shaw}},\ }\href {\doibase 10.1002/jcc.20267} {\bibfield  {journal} {\bibinfo
   {journal} {J. Comput. Chem.}\ }\textbf {\bibinfo {volume} {26}},\ \bibinfo
  {pages} {1318} (\bibinfo {year} {2005})}\BibitemShut {NoStop}%
\bibitem [{\citenamefont {Praprotnik}\ \emph
  {et~al.}(2005{\natexlab{a}})\citenamefont {Praprotnik}, \citenamefont
  {Site},\ and\ \citenamefont {Kremer}}]{praprotnik_adaptive_2005}%
  \BibitemOpen
  \bibfield  {author} {\bibinfo {author} {\bibfnamefont {M.}~\bibnamefont
  {Praprotnik}}, \bibinfo {author} {\bibfnamefont {L.~D.}\ \bibnamefont
  {Site}}, \ and\ \bibinfo {author} {\bibfnamefont {K.}~\bibnamefont
  {Kremer}},\ }\href {\doibase 10.1063/1.2132286} {\bibfield  {journal}
  {\bibinfo  {journal} {The Journal of Chemical Physics}\ }\textbf {\bibinfo
  {volume} {123}},\ \bibinfo {pages} {224106} (\bibinfo {year}
  {2005}{\natexlab{a}})}\BibitemShut {NoStop}%
\bibitem [{\citenamefont {Praprotnik}\ \emph {et~al.}(2007)\citenamefont
  {Praprotnik}, \citenamefont {Matysiak}, \citenamefont {Site}, \citenamefont
  {Kremer},\ and\ \citenamefont {Clementi}}]{praprotnik_adaptive_2007-1}%
  \BibitemOpen
  \bibfield  {author} {\bibinfo {author} {\bibfnamefont {M.}~\bibnamefont
  {Praprotnik}}, \bibinfo {author} {\bibfnamefont {S.}~\bibnamefont
  {Matysiak}}, \bibinfo {author} {\bibfnamefont {L.~D.}\ \bibnamefont {Site}},
  \bibinfo {author} {\bibfnamefont {K.}~\bibnamefont {Kremer}}, \ and\ \bibinfo
  {author} {\bibfnamefont {C.}~\bibnamefont {Clementi}},\ }\href
  {http://www.iop.org/EJ/abstract/0953-8984/19/29/292201/} {\bibfield
  {journal} {\bibinfo  {journal} {Journal of Physics: Condensed Matter}\
  }\textbf {\bibinfo {volume} {19}},\ \bibinfo {pages} {292201} (\bibinfo
  {year} {2007})}\BibitemShut {NoStop}%
\bibitem [{\citenamefont {Praprotnik}\ \emph {et~al.}(2009)\citenamefont
  {Praprotnik}, \citenamefont {Matysiak}, \citenamefont {Site}, \citenamefont
  {Kremer},\ and\ \citenamefont {Clementi}}]{praprotnik_corrigendum:_2009}%
  \BibitemOpen
  \bibfield  {author} {\bibinfo {author} {\bibfnamefont {M.}~\bibnamefont
  {Praprotnik}}, \bibinfo {author} {\bibfnamefont {S.}~\bibnamefont
  {Matysiak}}, \bibinfo {author} {\bibfnamefont {L.~D.}\ \bibnamefont {Site}},
  \bibinfo {author} {\bibfnamefont {K.}~\bibnamefont {Kremer}}, \ and\ \bibinfo
  {author} {\bibfnamefont {C.}~\bibnamefont {Clementi}},\ }\href
  {http://www.iop.org/EJ/abstract/0953-8984/21/49/499801/} {\bibfield
  {journal} {\bibinfo  {journal} {Journal of Physics: Condensed Matter}\
  }\textbf {\bibinfo {volume} {21}},\ \bibinfo {pages} {499801} (\bibinfo
  {year} {2009})}\BibitemShut {NoStop}%
\bibitem [{\citenamefont {Potestio}\ \emph
  {et~al.}(2013{\natexlab{b}})\citenamefont {Potestio}, \citenamefont
  {Espa{\~n}ol}, \citenamefont {Delgado-Buscalioni}, \citenamefont {Everaers},
  \citenamefont {Kremer},\ and\ \citenamefont {Donadio}}]{Potestio2013b}%
  \BibitemOpen
  \bibfield  {author} {\bibinfo {author} {\bibfnamefont {R.}~\bibnamefont
  {Potestio}}, \bibinfo {author} {\bibfnamefont {P.}~\bibnamefont
  {Espa{\~n}ol}}, \bibinfo {author} {\bibfnamefont {R.}~\bibnamefont
  {Delgado-Buscalioni}}, \bibinfo {author} {\bibfnamefont {R.}~\bibnamefont
  {Everaers}}, \bibinfo {author} {\bibfnamefont {K.}~\bibnamefont {Kremer}}, \
  and\ \bibinfo {author} {\bibfnamefont {D.}~\bibnamefont {Donadio}},\ }\href
  {\doibase 10.1103/PhysRevLett.111.060601} {\bibfield  {journal} {\bibinfo
  {journal} {Phys. Rev. Lett.}\ }\textbf {\bibinfo {volume} {111}},\ \bibinfo
  {pages} {060601} (\bibinfo {year} {2013}{\natexlab{b}})}\BibitemShut
  {NoStop}%
\bibitem [{\citenamefont {Kreis}\ \emph {et~al.}(2014)\citenamefont {Kreis},
  \citenamefont {Donadio}, \citenamefont {Kremer},\ and\ \citenamefont
  {Potestio}}]{Kreis2014}%
  \BibitemOpen
  \bibfield  {author} {\bibinfo {author} {\bibfnamefont {K.}~\bibnamefont
  {Kreis}}, \bibinfo {author} {\bibfnamefont {D.}~\bibnamefont {Donadio}},
  \bibinfo {author} {\bibfnamefont {K.}~\bibnamefont {Kremer}}, \ and\ \bibinfo
  {author} {\bibfnamefont {R.}~\bibnamefont {Potestio}},\ }\href {\doibase
  10.1209/0295-5075/108/30007} {\bibfield  {journal} {\bibinfo  {journal}
  {EPL}\ }\textbf {\bibinfo {volume} {108}},\ \bibinfo {pages} {30007}
  (\bibinfo {year} {2014})}\BibitemShut {NoStop}%
\bibitem [{\citenamefont {Fogarty}\ \emph {et~al.}(2015)\citenamefont
  {Fogarty}, \citenamefont {Potestio},\ and\ \citenamefont
  {Kremer}}]{Kremer_JCP_2015-adresprot}%
  \BibitemOpen
  \bibfield  {author} {\bibinfo {author} {\bibfnamefont {A.~C.}\ \bibnamefont
  {Fogarty}}, \bibinfo {author} {\bibfnamefont {R.}~\bibnamefont {Potestio}}, \
  and\ \bibinfo {author} {\bibfnamefont {K.}~\bibnamefont {Kremer}},\ }\href
  {\doibase 10.1063/1.4921347} {\bibfield  {journal} {\bibinfo  {journal} {J.
  Chem. Phys.}\ }\textbf {\bibinfo {volume} {142}},\ \bibinfo {eid} {195101}
  (\bibinfo {year} {2015})}\BibitemShut {NoStop}%
\bibitem [{\citenamefont {Kreis}\ \emph {et~al.}(2015)\citenamefont {Kreis},
  \citenamefont {Fogarty}, \citenamefont {Kremer},\ and\ \citenamefont
  {Potestio}}]{kreisEPJST2015}%
  \BibitemOpen
  \bibfield  {author} {\bibinfo {author} {\bibfnamefont {K.}~\bibnamefont
  {Kreis}}, \bibinfo {author} {\bibfnamefont {A.~C.}\ \bibnamefont {Fogarty}},
  \bibinfo {author} {\bibfnamefont {K.}~\bibnamefont {Kremer}}, \ and\ \bibinfo
  {author} {\bibfnamefont {R.}~\bibnamefont {Potestio}},\ }\href {\doibase
  10.1140/epjst/e2015-02412-1} {\bibfield  {journal} {\bibinfo  {journal} {Eur.
  Phys. J Special Topics}\ }\textbf {\bibinfo {volume} {224}},\ \bibinfo
  {pages} {2289} (\bibinfo {year} {2015})}\BibitemShut {NoStop}%
\bibitem [{\citenamefont {Zavadlav}\ \emph {et~al.}(2015)\citenamefont
  {Zavadlav}, \citenamefont {Podgornik},\ and\ \citenamefont
  {Praprotnik}}]{Praprotnik_JCTC_2015-dna}%
  \BibitemOpen
  \bibfield  {author} {\bibinfo {author} {\bibfnamefont {J.}~\bibnamefont
  {Zavadlav}}, \bibinfo {author} {\bibfnamefont {R.}~\bibnamefont {Podgornik}},
  \ and\ \bibinfo {author} {\bibfnamefont {M.}~\bibnamefont {Praprotnik}},\
  }\href {\doibase 10.1021/acs.jctc.5b00596} {\bibfield  {journal} {\bibinfo
  {journal} {J. Chem. Theory Comput.}\ }\textbf {\bibinfo {volume} {11}},\
  \bibinfo {pages} {5035} (\bibinfo {year} {2015})},\ \Eprint
  {http://arxiv.org/abs/http://dx.doi.org/10.1021/acs.jctc.5b00596}
  {http://dx.doi.org/10.1021/acs.jctc.5b00596} \BibitemShut {NoStop}%
\bibitem [{\citenamefont {Poma}\ and\ \citenamefont {{Delle
  Site}}(2010)}]{Poma2010}%
  \BibitemOpen
  \bibfield  {author} {\bibinfo {author} {\bibfnamefont {A.~B.}\ \bibnamefont
  {Poma}}\ and\ \bibinfo {author} {\bibfnamefont {L.}~\bibnamefont {{Delle
  Site}}},\ }\href {\doibase 10.1103/PhysRevLett.104.250201} {\bibfield
  {journal} {\bibinfo  {journal} {Phys. Rev. Lett.}\ }\textbf {\bibinfo
  {volume} {104}},\ \bibinfo {pages} {250201} (\bibinfo {year} {2010})},\
  \Eprint {http://arxiv.org/abs/1002.4118} {arXiv:1002.4118} \BibitemShut
  {NoStop}%
\bibitem [{\citenamefont {Poma}\ and\ \citenamefont {{Delle
  Site}}(2011)}]{Poma2011}%
  \BibitemOpen
  \bibfield  {author} {\bibinfo {author} {\bibfnamefont {A.~B.}\ \bibnamefont
  {Poma}}\ and\ \bibinfo {author} {\bibfnamefont {L.}~\bibnamefont {{Delle
  Site}}},\ }\href {\doibase 10.1039/c0cp02865g} {\bibfield  {journal}
  {\bibinfo  {journal} {Phys. Chem. Chem. Phys.}\ }\textbf {\bibinfo {volume}
  {13}},\ \bibinfo {pages} {10510} (\bibinfo {year} {2011})}\BibitemShut
  {NoStop}%
\bibitem [{\citenamefont {Kreis}\ \emph {et~al.}(0)\citenamefont {Kreis},
  \citenamefont {Tuckerman}, \citenamefont {Donadio}, \citenamefont {Kremer},\
  and\ \citenamefont {Potestio}}]{KreisJCTC2k16}%
  \BibitemOpen
  \bibfield  {author} {\bibinfo {author} {\bibfnamefont {K.}~\bibnamefont
  {Kreis}}, \bibinfo {author} {\bibfnamefont {M.~E.}\ \bibnamefont
  {Tuckerman}}, \bibinfo {author} {\bibfnamefont {D.}~\bibnamefont {Donadio}},
  \bibinfo {author} {\bibfnamefont {K.}~\bibnamefont {Kremer}}, \ and\ \bibinfo
  {author} {\bibfnamefont {R.}~\bibnamefont {Potestio}},\ }\href {\doibase
  10.1021/acs.jctc.6b00242} {\bibfield  {journal} {\bibinfo  {journal} {Journal
  of Chemical Theory and Computation}\ }\textbf {\bibinfo {volume} {0}},\
  \bibinfo {pages} {null} (\bibinfo {year} {0})},\ \bibinfo {note} {pMID:
  27214610},\ \Eprint
  {http://arxiv.org/abs/http://dx.doi.org/10.1021/acs.jctc.6b00242}
  {http://dx.doi.org/10.1021/acs.jctc.6b00242} \BibitemShut {NoStop}%
\bibitem [{\citenamefont {Zavadlav}\ \emph {et~al.}(2014)\citenamefont
  {Zavadlav}, \citenamefont {Melo}, \citenamefont {Marrink},\ and\
  \citenamefont {Praprotnik}}]{Praprotnik_JChemPhys_2014-adres_bio}%
  \BibitemOpen
  \bibfield  {author} {\bibinfo {author} {\bibfnamefont {J.}~\bibnamefont
  {Zavadlav}}, \bibinfo {author} {\bibfnamefont {M.~N.}\ \bibnamefont {Melo}},
  \bibinfo {author} {\bibfnamefont {S.~J.}\ \bibnamefont {Marrink}}, \ and\
  \bibinfo {author} {\bibfnamefont {M.}~\bibnamefont {Praprotnik}},\ }\href
  {\doibase 10.1063/1.4863329} {\bibfield  {journal} {\bibinfo  {journal} {J.
  Chem. Phys.}\ }\textbf {\bibinfo {volume} {140}},\ \bibinfo {eid} {054114}
  (\bibinfo {year} {2014})}\BibitemShut {NoStop}%
\bibitem [{\citenamefont {Praprotnik}\ \emph
  {et~al.}(2005{\natexlab{b}})\citenamefont {Praprotnik}, \citenamefont {{Delle
  Site}},\ and\ \citenamefont {Kremer}}]{Kremer_JCP_2005-onthefly}%
  \BibitemOpen
  \bibfield  {author} {\bibinfo {author} {\bibfnamefont {M.}~\bibnamefont
  {Praprotnik}}, \bibinfo {author} {\bibfnamefont {L.}~\bibnamefont {{Delle
  Site}}}, \ and\ \bibinfo {author} {\bibfnamefont {K.}~\bibnamefont
  {Kremer}},\ }\href {\doibase 10.1063/1.2132286} {\bibfield  {journal}
  {\bibinfo  {journal} {J. Chem. Phys.}\ }\textbf {\bibinfo {volume} {123}},\
  \bibinfo {pages} {224106} (\bibinfo {year} {2005}{\natexlab{b}})}\BibitemShut
  {NoStop}%
\bibitem [{\citenamefont {Praprotnik}\ \emph
  {et~al.}(2008{\natexlab{b}})\citenamefont {Praprotnik}, \citenamefont
  {Site},\ and\ \citenamefont {Kremer}}]{Praprotnik2008review}%
  \BibitemOpen
  \bibfield  {author} {\bibinfo {author} {\bibfnamefont {M.}~\bibnamefont
  {Praprotnik}}, \bibinfo {author} {\bibfnamefont {L.~D.}\ \bibnamefont
  {Site}}, \ and\ \bibinfo {author} {\bibfnamefont {K.}~\bibnamefont
  {Kremer}},\ }\href {\doibase 10.1146/annurev.physchem.59.032607.093707}
  {\bibfield  {journal} {\bibinfo  {journal} {Annu. Rev. Phys. Chem.}\ }\textbf
  {\bibinfo {volume} {59}},\ \bibinfo {pages} {545} (\bibinfo {year}
  {2008}{\natexlab{b}})},\ \Eprint {http://arxiv.org/abs/0510223v1}
  {arXiv:0510223v1 [arXiv:cond-mat]} \BibitemShut {NoStop}%
\bibitem [{\citenamefont {Hess}\ \emph {et~al.}(2008)\citenamefont {Hess},
  \citenamefont {Kutzner}, \citenamefont {van~der Spoel},\ and\ \citenamefont
  {Lindahl}}]{gromacs4}%
  \BibitemOpen
  \bibfield  {author} {\bibinfo {author} {\bibfnamefont {B.}~\bibnamefont
  {Hess}}, \bibinfo {author} {\bibfnamefont {C.}~\bibnamefont {Kutzner}},
  \bibinfo {author} {\bibfnamefont {D.}~\bibnamefont {van~der Spoel}}, \ and\
  \bibinfo {author} {\bibfnamefont {E.}~\bibnamefont {Lindahl}},\ }\href@noop
  {} {\bibfield  {journal} {\bibinfo  {journal} {J. Chem. Theory Comput.}\
  }\textbf {\bibinfo {volume} {4}},\ \bibinfo {pages} {435} (\bibinfo {year}
  {2008})}\BibitemShut {NoStop}%
\bibitem [{\citenamefont {Flynn}\ and\ \citenamefont
  {Hummel}(1992)}]{FlynnDLB1992}%
  \BibitemOpen
  \bibfield  {author} {\bibinfo {author} {\bibfnamefont {L.}~\bibnamefont
  {Flynn}}\ and\ \bibinfo {author} {\bibfnamefont {S.}~\bibnamefont {Hummel}},\
  }\href@noop {} {\bibfield  {journal} {\bibinfo  {journal} {IBM Research
  Report}\ }\textbf {\bibinfo {volume} {RC18462}} (\bibinfo {year}
  {1992})}\BibitemShut {NoStop}%
\bibitem [{\citenamefont {Hummel}\ \emph {et~al.}(1991)\citenamefont {Hummel},
  \citenamefont {Schonberg},\ and\ \citenamefont {Flynn}}]{HummelSC1991}%
  \BibitemOpen
  \bibfield  {author} {\bibinfo {author} {\bibfnamefont {S.~F.}\ \bibnamefont
  {Hummel}}, \bibinfo {author} {\bibfnamefont {E.}~\bibnamefont {Schonberg}}, \
  and\ \bibinfo {author} {\bibfnamefont {L.~E.}\ \bibnamefont {Flynn}},\ }in\
  \href {\doibase 10.1145/125826.126137} {\emph {\bibinfo {booktitle}
  {{Proceedings of the 1991 ACM/IEEE Conference on Supercomputing}}}},\
  \bibinfo {series and number} {{Supercomputing '91}}\ (\bibinfo  {publisher}
  {ACM},\ \bibinfo {address} {New York, NY, USA},\ \bibinfo {year} {1991})\
  pp.\ \bibinfo {pages} {610--632}\BibitemShut {NoStop}%
\bibitem [{\citenamefont {Polychronopoulos}\ and\ \citenamefont
  {Kuck}(1987)}]{PolyITC1987}%
  \BibitemOpen
  \bibfield  {author} {\bibinfo {author} {\bibfnamefont {C.~D.}\ \bibnamefont
  {Polychronopoulos}}\ and\ \bibinfo {author} {\bibfnamefont {D.~J.}\
  \bibnamefont {Kuck}},\ }\href {\doibase 10.1109/TC.1987.5009495} {\bibfield
  {journal} {\bibinfo  {journal} {IEEE Trans. Comput.}\ }\textbf {\bibinfo
  {volume} {36}},\ \bibinfo {pages} {1425} (\bibinfo {year}
  {1987})}\BibitemShut {NoStop}%
\bibitem [{Blu()}]{BlumofeICS1994}%
  \BibitemOpen
  \bibfield  {booktitle} {\emph {\bibinfo {booktitle} {{Proceedings of the 35th
  Annual Symposium on Foundations of Computer Science}}},\ }\href@noop {} {\
  \bibinfo {series} {{SFCS '94}}}\BibitemShut {NoStop}%
\bibitem [{\citenamefont {Kruskal}\ and\ \citenamefont
  {Weiss}(1985)}]{KruskalPP1985}%
  \BibitemOpen
  \bibfield  {author} {\bibinfo {author} {\bibfnamefont {C.~P.}\ \bibnamefont
  {Kruskal}}\ and\ \bibinfo {author} {\bibfnamefont {A.}~\bibnamefont
  {Weiss}},\ }\href {\doibase 10.1109/TSE.1985.231547} {\bibfield  {journal}
  {\bibinfo  {journal} {IEEE Transactions on Software Engineering}\ }\textbf
  {\bibinfo {volume} {SE-11}},\ \bibinfo {pages} {1001} (\bibinfo {year}
  {1985})}\BibitemShut {NoStop}%
\bibitem [{\citenamefont {Rudolph}\ \emph {et~al.}(1991)\citenamefont
  {Rudolph}, \citenamefont {Slivkin-Allalouf},\ and\ \citenamefont
  {Upfal}}]{RudolphSPAA1991}%
  \BibitemOpen
  \bibfield  {author} {\bibinfo {author} {\bibfnamefont {L.}~\bibnamefont
  {Rudolph}}, \bibinfo {author} {\bibfnamefont {M.}~\bibnamefont
  {Slivkin-Allalouf}}, \ and\ \bibinfo {author} {\bibfnamefont
  {E.}~\bibnamefont {Upfal}},\ }in\ \href {\doibase 10.1145/113379.113401}
  {\emph {\bibinfo {booktitle} {{Proceedings of the Third Annual ACM Symposium
  on Parallel Algorithms and Architectures}}}},\ \bibinfo {series and number}
  {{SPAA '91}}\ (\bibinfo  {publisher} {ACM},\ \bibinfo {address} {New York,
  NY, USA},\ \bibinfo {year} {1991})\ pp.\ \bibinfo {pages}
  {237--245}\BibitemShut {NoStop}%
\bibitem [{\citenamefont {Kale}\ and\ \citenamefont {Zheng}(2009)}]{CharmPP}%
  \BibitemOpen
  \bibfield  {author} {\bibinfo {author} {\bibfnamefont {L.~V.}\ \bibnamefont
  {Kale}}\ and\ \bibinfo {author} {\bibfnamefont {G.}~\bibnamefont {Zheng}},\
  }in\ \href@noop {} {\emph {\bibinfo {booktitle} {{Advanced Computational
  Infrastructures for Parallel and Distributed Applications}}}},\ \bibinfo
  {editor} {edited by\ \bibinfo {editor} {\bibfnamefont {M.}~\bibnamefont
  {Parashar}}}\ (\bibinfo  {publisher} {Wiley-Interscience},\ \bibinfo {year}
  {2009})\ pp.\ \bibinfo {pages} {265--282}\BibitemShut {NoStop}%
\bibitem [{\citenamefont {Kaiser}\ \emph {et~al.}(2014)\citenamefont {Kaiser},
  \citenamefont {Heller}, \citenamefont {Adelstein-Lelbach}, \citenamefont
  {Serio},\ and\ \citenamefont {Fey}}]{HPX}%
  \BibitemOpen
  \bibfield  {author} {\bibinfo {author} {\bibfnamefont {H.}~\bibnamefont
  {Kaiser}}, \bibinfo {author} {\bibfnamefont {T.}~\bibnamefont {Heller}},
  \bibinfo {author} {\bibfnamefont {B.}~\bibnamefont {Adelstein-Lelbach}},
  \bibinfo {author} {\bibfnamefont {A.}~\bibnamefont {Serio}}, \ and\ \bibinfo
  {author} {\bibfnamefont {D.}~\bibnamefont {Fey}},\ }in\ \href {\doibase
  10.1145/2676870.2676883} {\emph {\bibinfo {booktitle} {{Proceedings of the
  8th International Conference on Partitioned Global Address Space Programming
  Models}}}},\ \bibinfo {series and number} {{PGAS '14}}\ (\bibinfo
  {publisher} {ACM},\ \bibinfo {address} {New York, NY, USA},\ \bibinfo {year}
  {2014})\ pp.\ \bibinfo {pages} {6:1--6:11}\BibitemShut {NoStop}%
\bibitem [{\citenamefont {Chapin}\ \emph {et~al.}(1999)\citenamefont {Chapin},
  \citenamefont {Katramatos}, \citenamefont {Karpovich},\ and\ \citenamefont
  {Grimshaw}}]{Legion}%
  \BibitemOpen
  \bibfield  {author} {\bibinfo {author} {\bibfnamefont {S.~J.}\ \bibnamefont
  {Chapin}}, \bibinfo {author} {\bibfnamefont {D.}~\bibnamefont {Katramatos}},
  \bibinfo {author} {\bibfnamefont {J.}~\bibnamefont {Karpovich}}, \ and\
  \bibinfo {author} {\bibfnamefont {A.~S.}\ \bibnamefont {Grimshaw}},\
  }\enquote {\bibinfo {title} {{The Legion Resource Management System}},}\ in\
  \href {\doibase 10.1007/3-540-47954-6_9} {\emph {\bibinfo {booktitle} {{Job
  Scheduling Strategies for Parallel Processing: IPPS/SPDP'99Workshop, JSSPP'99
  San Juan, Puerto Rico, April 16, 1999 Proceedings}}}},\ \bibinfo {editor}
  {edited by\ \bibinfo {editor} {\bibfnamefont {D.~G.}\ \bibnamefont
  {Feitelson}}\ and\ \bibinfo {editor} {\bibfnamefont {L.}~\bibnamefont
  {Rudolph}}}\ (\bibinfo  {publisher} {Springer Berlin Heidelberg},\ \bibinfo
  {address} {Berlin, Heidelberg},\ \bibinfo {year} {1999})\ pp.\ \bibinfo
  {pages} {162--178}\BibitemShut {NoStop}%
\bibitem [{\citenamefont {Allen}\ and\ \citenamefont
  {Tildesley}(1987)}]{allen87}%
  \BibitemOpen
  \bibfield  {author} {\bibinfo {author} {\bibfnamefont {M.~P.}\ \bibnamefont
  {Allen}}\ and\ \bibinfo {author} {\bibfnamefont {D.~J.}\ \bibnamefont
  {Tildesley}},\ }\href@noop {} {\emph {\bibinfo {title} {{Computer Simulation
  of Liquids}}}}\ (\bibinfo  {publisher} {Oxford University Press},\ \bibinfo
  {address} {New York},\ \bibinfo {year} {1987})\BibitemShut {NoStop}%
\bibitem [{\citenamefont {Osprey}\ \emph {et~al.}(2014)\citenamefont {Osprey},
  \citenamefont {Riley}, \citenamefont {Manjunathaiah},\ and\ \citenamefont
  {Lawrence}}]{OspreyHPCS2014}%
  \BibitemOpen
  \bibfield  {author} {\bibinfo {author} {\bibfnamefont {A.}~\bibnamefont
  {Osprey}}, \bibinfo {author} {\bibfnamefont {G.~D.}\ \bibnamefont {Riley}},
  \bibinfo {author} {\bibfnamefont {M.}~\bibnamefont {Manjunathaiah}}, \ and\
  \bibinfo {author} {\bibfnamefont {B.~N.}\ \bibnamefont {Lawrence}},\ }in\
  \href {http://dblp.uni-trier.de/db/conf/hpcs/hpcs2014.html#OspreyRML14;
  http://dx.doi.org/10.1109/HPCSim.2014.6903760} {\emph {\bibinfo {booktitle}
  {{HPCS}}}}\ (\bibinfo  {publisher} {IEEE},\ \bibinfo {year} {2014})\ pp.\
  \bibinfo {pages} {715--723}\BibitemShut {NoStop}%
\bibitem [{\citenamefont {Malik}\ \emph {et~al.}(2016)\citenamefont {Malik},
  \citenamefont {Rychkov},\ and\ \citenamefont {Lastovetsky}}]{MalikCCPE2016}%
  \BibitemOpen
  \bibfield  {author} {\bibinfo {author} {\bibfnamefont {T.}~\bibnamefont
  {Malik}}, \bibinfo {author} {\bibfnamefont {V.}~\bibnamefont {Rychkov}}, \
  and\ \bibinfo {author} {\bibfnamefont {A.~L.}\ \bibnamefont {Lastovetsky}},\
  }\href
  {http://dblp.uni-trier.de/db/journals/concurrency/concurrency28.html#MalikRL16;
  http://dx.doi.org/10.1002/cpe.3609} {\bibfield  {journal} {\bibinfo
  {journal} {Concurrency and Computation: Practice and Experience}\ }\textbf
  {\bibinfo {volume} {28}},\ \bibinfo {pages} {802} (\bibinfo {year}
  {2016})}\BibitemShut {NoStop}%
\bibitem [{\citenamefont {Markidis}\ \emph {et~al.}(2015)\citenamefont
  {Markidis}, \citenamefont {Vencels}, \citenamefont {Peng}, \citenamefont
  {Akhmetova}, \citenamefont {Laure},\ and\ \citenamefont
  {Henri}}]{HenriPRE2015}%
  \BibitemOpen
  \bibfield  {author} {\bibinfo {author} {\bibfnamefont {S.}~\bibnamefont
  {Markidis}}, \bibinfo {author} {\bibfnamefont {J.}~\bibnamefont {Vencels}},
  \bibinfo {author} {\bibfnamefont {I.~B.}\ \bibnamefont {Peng}}, \bibinfo
  {author} {\bibfnamefont {D.}~\bibnamefont {Akhmetova}}, \bibinfo {author}
  {\bibfnamefont {E.}~\bibnamefont {Laure}}, \ and\ \bibinfo {author}
  {\bibfnamefont {P.}~\bibnamefont {Henri}},\ }\href {\doibase
  10.1103/PhysRevE.91.013306} {\bibfield  {journal} {\bibinfo  {journal} {Phys.
  Rev. E}\ }\textbf {\bibinfo {volume} {91}},\ \bibinfo {pages} {013306}
  (\bibinfo {year} {2015})}\BibitemShut {NoStop}%
\bibitem [{GHH()}]{GHH}%
  \BibitemOpen
  \href@noop {} {\enquote {\bibinfo {title}
  {{https://github.com/govarguz/HeSpaDDA.git}},}\ }\BibitemShut {NoStop}%
\bibitem [{\citenamefont {Soper}(1996)}]{Soper_JCP_1996}%
  \BibitemOpen
  \bibfield  {author} {\bibinfo {author} {\bibfnamefont {A.}~\bibnamefont
  {Soper}},\ }\href {\doibase 10.1016/0301-0104(95)00357-6} {\bibfield
  {journal} {\bibinfo  {journal} {Chemical Physics}\ }\textbf {\bibinfo
  {volume} {202}},\ \bibinfo {pages} {295–306} (\bibinfo {year}
  {1996})}\BibitemShut {NoStop}%
\bibitem [{\citenamefont {Tschöp}\ \emph {et~al.}(1998)\citenamefont
  {Tschöp}, \citenamefont {Kremer}, \citenamefont {Batoulis}, \citenamefont
  {Bürger},\ and\ \citenamefont {Hahn}}]{KK_AcPoly_1998}%
  \BibitemOpen
  \bibfield  {author} {\bibinfo {author} {\bibfnamefont {W.}~\bibnamefont
  {Tschöp}}, \bibinfo {author} {\bibfnamefont {K.}~\bibnamefont {Kremer}},
  \bibinfo {author} {\bibfnamefont {J.}~\bibnamefont {Batoulis}}, \bibinfo
  {author} {\bibfnamefont {T.}~\bibnamefont {Bürger}}, \ and\ \bibinfo
  {author} {\bibfnamefont {O.}~\bibnamefont {Hahn}},\ }\href {\doibase
  10.1002/(SICI)1521-4044(199802)49:2/3<61::AID-APOL61>3.0.CO;2-V} {\bibfield
  {journal} {\bibinfo  {journal} {Acta Polymerica}\ }\textbf {\bibinfo {volume}
  {49}},\ \bibinfo {pages} {61} (\bibinfo {year} {1998})}\BibitemShut {NoStop}%
\bibitem [{\citenamefont {Reith}\ \emph {et~al.}(2003)\citenamefont {Reith},
  \citenamefont {P{\"u}tz},\ and\ \citenamefont
  {M{\"u}ller-Plathe}}]{Reith2003}%
  \BibitemOpen
  \bibfield  {author} {\bibinfo {author} {\bibfnamefont {D.}~\bibnamefont
  {Reith}}, \bibinfo {author} {\bibfnamefont {M.}~\bibnamefont {P{\"u}tz}}, \
  and\ \bibinfo {author} {\bibfnamefont {F.}~\bibnamefont
  {M{\"u}ller-Plathe}},\ }\href {\doibase 10.1002/jcc.10307} {\bibfield
  {journal} {\bibinfo  {journal} {J. Comput. Chem.}\ }\textbf {\bibinfo
  {volume} {24}},\ \bibinfo {pages} {1624} (\bibinfo {year} {2003})},\ \Eprint
  {http://arxiv.org/abs/0211454} {arXiv:0211454 [cond-mat]} \BibitemShut
  {NoStop}%
\bibitem [{\citenamefont {Páll}\ \emph {et~al.}(2015)\citenamefont {Páll},
  \citenamefont {Abraham}, \citenamefont {Kutzner}, \citenamefont {Hess},\ and\
  \citenamefont {Lindahl}}]{Gromacs2015}%
  \BibitemOpen
  \bibfield  {author} {\bibinfo {author} {\bibfnamefont {S.}~\bibnamefont
  {Páll}}, \bibinfo {author} {\bibfnamefont {M.~J.}\ \bibnamefont {Abraham}},
  \bibinfo {author} {\bibfnamefont {C.}~\bibnamefont {Kutzner}}, \bibinfo
  {author} {\bibfnamefont {B.}~\bibnamefont {Hess}}, \ and\ \bibinfo {author}
  {\bibfnamefont {E.}~\bibnamefont {Lindahl}},\ }\enquote {\bibinfo {title}
  {{Tackling Exascale Software Challenges in Molecular Dynamics Simulations
  with GROMACS}},}\ in\ \href {\doibase 10.1007/978-3-319-15976-8_1} {\emph
  {\bibinfo {booktitle} {{Solving Software Challenges for Exascale:
  International Conference on Exascale Applications and Software, EASC 2014,
  Stockholm, Sweden, April 2-3, 2014, Revised Selected Papers}}}},\ \bibinfo
  {editor} {edited by\ \bibinfo {editor} {\bibfnamefont {S.}~\bibnamefont
  {Markidis}}\ and\ \bibinfo {editor} {\bibfnamefont {E.}~\bibnamefont
  {Laure}}}\ (\bibinfo  {publisher} {Springer International Publishing},\
  \bibinfo {address} {Cham},\ \bibinfo {year} {2015})\ p.\ \bibinfo {pages}
  {3–27}\BibitemShut {NoStop}%
\end{thebibliography}%
\end{document}